\RequirePackage{fix-cm}
\documentclass[smallextended]{svjour3}       
\smartqed  
\usepackage{graphicx}
\usepackage{url}
\usepackage{graphicx}
\usepackage{comment}
\usepackage{multirow}
\usepackage{color}
\usepackage{lscape}
\usepackage{longtable}
\usepackage{supertabular}
\usepackage{natbib}
\usepackage{enumitem}
\usepackage{changepage}
\usepackage{tabularx}
\usepackage[table]{xcolor}
\usepackage[nomessages]{fp}
\usepackage{pgfplots}
\usepackage{venndiagram}
\usepackage{listings}
\usepackage{pgfplotstable}
\usepackage[flushleft]{threeparttable}
\usepackage{caption}
\usepackage{subfig}
\usepackage{booktabs}

\newlength{\maxlen}

\newcommand{\RqOne}{\textbf{RQ$_1$:} \emph{Can the values of \texttt{diff}-related metrics become different because of different \texttt{diff} algorithms?}}
\newcommand{\RqTwo}{\textbf{RQ$_2$:} \emph{Are the results of bug-introducing change identification different because of different \texttt{diff} algorithms?}}
\newcommand{\RqThree}{\textbf{RQ$_3$:} \emph{Which \texttt{diff} algorithm is better in generating a good \texttt{diff}?}}

\newcommand{\dataset}{\url{https://github.com/yusufsn/DifferentDiffAlgorithms}}

\begin{document}

\title{How Different Are Different \textit{diff} Algorithms in Git?}
\subtitle{Use \texttt{--histogram} for Code Changes}


\author{Yusuf Sulistyo Nugroho  \and
        Hideaki Hata	\and
        Kenichi Matsumoto
}


\institute{Yusuf Sulistyo Nugroho, Hideaki Hata,
              Kenichi Matsumoto \at
              Nara Institute of Science and Technology \\
              \email{\{yusuf\_sulistyo.nugroho.yi5, hata,
              matumoto\}@is.naist.jp}    \\ 
}

\date{Received: date / Accepted: date}

\maketitle

\begin{abstract}
Automatic identification of the differences between two versions of a file is a common and basic task in several applications of mining code repositories.
Git, a version control system, has a diff utility and users can select algorithms of diff from the default algorithm \textit{Myers} to the advanced \textit{Histogram} algorithm.
From our systematic mapping, we identified three popular applications of diff in recent studies.
On the impact on code churn metrics in 14 Java projects, we obtained different values in 1.7\% to 8.2\% commits based on the different diff algorithms.
Regarding bug-introducing change identification, we found 6.0\% and 13.3\% in the identified bug-fix commits had different results of bug-introducing changes from 10 Java projects.
For patch application, we found that the \textit{Histogram} is more suitable than \textit{Myers} for providing the changes of code, from our manual analysis.
Thus, we strongly recommend using the \textit{Histogram} algorithm when mining Git repositories to consider differences in source code.

\end{abstract}

\section{Introduction}
\label{intro}
The \texttt{diff} utility calculates and displays the differences between two files, and is typically used to investigate the changes between two versions of the same file. Since understanding and measuring changes in software artifact is essential in empirical software engineering research, \texttt{diff} is commonly used in various topics, such as defect prediction where code churn~\citep{Nagappan:2005:URC:1062455.1062514,Shin:2011:ECC:2086332.2086581} and process metrics~\citep{Hata:2012:BPB:2337223.2337247,Madeyski:2015:PMS:2767937.2767961,7476771} are used,
code authorship~\citep{Rahman:2011:OED:1985793.1985860,6676896}, 
clone genealogy~\citep{Kim:2005:ESC:1081706.1081737,Duala-Ekoko:2007:TCC:1248820.1248849}, and
empirical studies of changes~\citep{Barr:2014:PSH:2635868.2635898,Ray:2015:UCC:2820518.2820526}.

Along with the growth of GitHub, recent studies analyze software changes from Git repositories by using the \texttt{git} command. 
Git, a version control system, offers \texttt{diff} utility for users to select the algorithms of \texttt{diff}.
Git offers four \texttt{diff} algorithms, namely, \textit{Myers}, \textit{Minimal}, \textit{Patience}, and \textit{Histogram}. 
Without an identifying algorithm, \textit{Myers} is used as the default algorithm.

In textual differencing, all \texttt{diff} algorithms are computationally correct in generating the \texttt{diff} outputs.
However, the \texttt{diff} outputs are sometimes different due to different \texttt{diff} algorithms.
Different \texttt{diff} algorithms might identify different change hunks, that is, a list of program statements deleted or added contiguously, separated by at least one line of unchanged context~\citep{Ray:2015:UCC:2820518.2820526}.
We expect that a set of changing operations done by developers can be represented by change hunks.
However, there can be inappropriate identifications of change hunks.
Although \textit{Histogram} that was introduced in \texttt{git 1.7.7}\footnote{\label{git177}\url{https://github.com/git/git/blob/77bd3ea9f54f1584147b594abc04c26ca516d987/Documentation/RelNotes/1.7.7.txt#L68-L70}} in 2011 might give better performance to \texttt{git diff}, it is not popular among software engineer communities.
Thus, we focus on the \textit{Myers} and \textit{Histogram} algorithms to empirically investigate the impact on software engineering research.
The motivation of this study is try to clarify the impact of adopting different \texttt{diff} algorithms on empirical studies and investigate which \texttt{diff} algorithm can provide better \texttt{diff} results that can be expected to recover the changing operations.
Furthermore, our study provides a comprehensive procedure of \textit{Myers} and \textit{Histogram} in generating the \texttt{diff}s and shows the differences between their outputs.
To the best of our knowledge, empirical comparisons of different \texttt{diff} algorithms in \texttt{git diff} command have never been undertaken. 
In this paper, we carry out two sequential analyses: systematic mapping and empirical comparisons. 

For the systematic mapping, we collect papers from three high ranking journals and eight top international conference proceedings published from 2013 to 2017. 
We then map 52 identified papers in the following four aspects: frequency of \texttt{diff} algorithms, analyzed software artifact, purpose of mining Git repositories, and data origins.
The results of the systematic mapping revealed that the advanced \texttt{diff} algorithms had not been considered in the previous studies. 
In terms of the focus of the \texttt{git} command, 51 out of 52 papers centralized on mining the code changes. 
We also found that the purposes of using the \texttt{git} command were to get patches (46.2\%), followed by metrics collection (25\%), and bug-introducing change identification (SZZ algorithm) (23.1\%). 
Regarding the dataset, most papers investigated OSS projects (98\%), even though the remaining work analyzed industrial data.

In our empirical analyses, we conduct three comparisons based on the most popular usages of \texttt{git diff} found in our mapping study: collecting metrics, identifying bug introduction, and getting patches. 
We investigate the disagreement between two \texttt{diff} algorithms: \textit{Myers} and \textit{Histogram}, and take a manual measurement of their quality in generating the \texttt{diff} lists. 
Based on previous related studies, we investigate the code changes from the files in 14 OSS projects that employ Continuous Integration for metrics collection and 10 Apache projects for the bug introduction identification to quantify the differences of the \texttt{diff} outputs that resulted from both \texttt{diff} algorithms. 
We analyze the quality of patches derived from \textit{Myers} and \textit{Histogram} by manually comparing their two \texttt{diff} from 377 changes, a statistically representative sample of the 21,590 changes identified in the above two comparisons. 
Our findings show that using various \texttt{diff} algorithms in the \texttt{git diff} command produced unequal \texttt{diff} lists. 

This influences the different number of files that have dissimilar added and deleted lines of code in each CI-Java project. 
The differences of these added and deleted lines that are distinguished by their different number and position range from 0.8\% to 6.2\% and from 1.4\% to 7.6\%, respectively.
The divergent \texttt{diff} outputs also affected the different number of identified files in bug introduction identification. 
The percentage of files that have different deleted lines of code range from 2.4\% to 6.6\%. 
Regarding the result of the patches analysis, we found that, in-code changes,
\textit{Histogram} is better in 62.6\% files,
while \textit{Myers} is better in 16.9\% files.
However, both \texttt{diff} algorithms evenly have a good quality in generating the list of non-code changes.

In sum, the contributions of this work are:
\begin{itemize}
    \item A systematic survey of studies that use \texttt{diff};    
    \item An analysis of metrics collected from \texttt{diff} outputs produced by \textit{Myers} and \textit{Histogram};
    \item An analysis of \textit{Myers} and \textit{Histogram} outputs in identifying potential bug-introducing changes;
    \item A manual comparison between \textit{Myers} and \textit{Histogram} to investigate their output quality.
\end{itemize}

The remaining parts of this paper are structured as follows. 
Section \ref{sec:codediffering} presents the application of various category of \texttt{diff} algorithms in the literature.
Section \ref{sec:algorithm} presents a brief explanation of \texttt{diff} algorithms used in the \texttt{git} command. 
We explain the differences between two \texttt{diff} algorithms in generating the list of changes. 
Section \ref{sec:survey} describes how we conduct a systematic mapping study and present the result of the survey. 
The overview of the three comparisons and the research questions are presented in Section~\ref{sec:overviewandRQ}.
Sections \ref{sec:metric}, \ref{sec:szz} and \ref{sec:patch} report our procedures and discuss their results in performing three comparison studies; namely, collecting metrics, identifying bug introduction, and getting patches respectively. 
In Section \ref{sec:discussion}, we discuss the implication of different \texttt{diff} algorithms and provide the example, and discuss their threats to validity, and finally we conclude in Section \ref{sec:conclusion}.

We have provided the data sets used in this paper publicly on the Web\footnote{\dataset}.

\section{Source Code Differencing}
\label{sec:codediffering}
Existing differencing techniques use similarities in names and structure to match code elements at a particular granularity, such as text-based and abstract-syntax-tree-based (AST).

Tree-based differencing techniques are widely used nowadays (e.g., diff in Unix), since they are expected to have better understandability than the text-based.
Such AST differencing tools were used in several studies. 
For example, Change Distilling (CD) that extracts the code changes by finding both a match between the nodes of the compared two abstract syntax trees and a minimum edit script that can transform one tree into the other given the computed matching~\citep{Fluri:2007:CDT:1314036.1314081}.
In this study, the text-based differencing is used to extract the changes at the beginning of the process as the input before further processed using the proposed AST algorithm.
In comparison with textual \texttt{diff}, the Change Distiller is able to assign the type of the changes such as declaration or body part of a method, rather than just to a line number.
Diff/TS~\citep{Hashimoto:2008:4656419} and MTDIFF~\citep{Dotzler:7582801} use moving code to compute the changes.
Diff/TS is used to analyze fine-grained structural change between versions of programs but only capable of processing \textit{Python, Java, C}, and \textit{C++} projects, while MTDIFF improves the accuracy of the previous tree-based approaches in detecting moved code.
~\cite{Falleri:2014:FAS:2642937.2642982} introduced an algorithm to compute edit scripts at the abstract syntax tree granularity including move actions.
In this study, the authors conducted a performance study to measure the running time and memory consumption between their proposed algorithm and the other tools, such as \texttt{GumTree} and \texttt{RTED} algorithm.
The classical text \texttt{diff} was used to present the reference values when comparing the running time between the involved algorithms.
Tree-based differencing approach was also used by~\cite{Higo:2017:GSA:3155562.3155630} to consider copy-and-paste as a type of editing action forming tree-based edit script, and~\cite{Huang:2018:CGC:3238147.3238219} to propose CLDIFF for generating concise linked code differences whose granularity is in between the existing code differencing and code change summarization methods.

Despite many advantages in tree-based differencing techniques,  text-based \texttt{diff} is widely used for several applications in software engineering research because of its simplicity and lightweight runtime.
Therefore, in this paper we only focus on studying the impact of changing \texttt{diff} algorithms, instead of comparing wider categories of differencing techniques.

\section{Diff Algorithms in Git}
\label{sec:algorithm}
\texttt{Diff} is an automatic comparison program used to find the disagreements between the older and the newer version of the same file in a storage (including insertions, deletions, document renaming, document movements etc.). 
The \texttt{diff} utility extracts code changes line by line in one file compared to the other file and reports them in a list. 
The operation of the \texttt{diff} program has been fundamentally solved by using the longest common subsequence (LCS) problem initiated by~\cite{Hunt75analgorithm}.
Since its first run on the Unix operating system in 1970, the \texttt{diff} command has been widely used in many studies.

The \texttt{git diff} command has numerous options in the application of code changes extraction\footnote{\label{git-difscm}\url{https://git-scm.com/docs/git-diff}}, including extracting changes related to the index and commit, paths on a filesystem, the original contents of objects, or even quantifying the number of changes for each object relatively from the sources. Researchers and practitioners are able to use the variation of these available options depending on their needs in extracting the data, not to mention, the \texttt{diff} algorithms.
The essence of \texttt{diff} algorithms is in contrasting the two sequences and to receive insight of the transformation from the first into the second by a series of operations using the ordered deletion and insertion. 
The subsequence can be flagged as a change if a delete and an insert concur on the same scope. 
The \texttt{diff} algorithm can be selected with this option \texttt{--diff-algorithm=<algorithm>}.

In Git, there are four \texttt{diff} algorithms, namely \textit{Myers}, \textit{Minimal}, \textit{Patience}, and \textit{Histogram}, which are utilized to obtain the differences of the two same files located in two different commits. 
The \textit{Minimal} and the \textit{Histogram} algorithms are the improved versions of the \textit{Myers} and the \textit{Patience} respectively. 
Each algorithm has its own procedures for finding the items presented in the original document, but absent in the second one and vice versa; as a consequence, different outputs may be produced. 
Due to the similarity of the basic idea of \textit{Minimal} and \textit{Histogram} algorithms with their precursors, in this paper we only contrasted the two \texttt{diff} algorithms: \textit{Myers} and \textit{Histogram}.

\begin{figure}[]
    \center
        \includegraphics[width=0.9\textwidth]{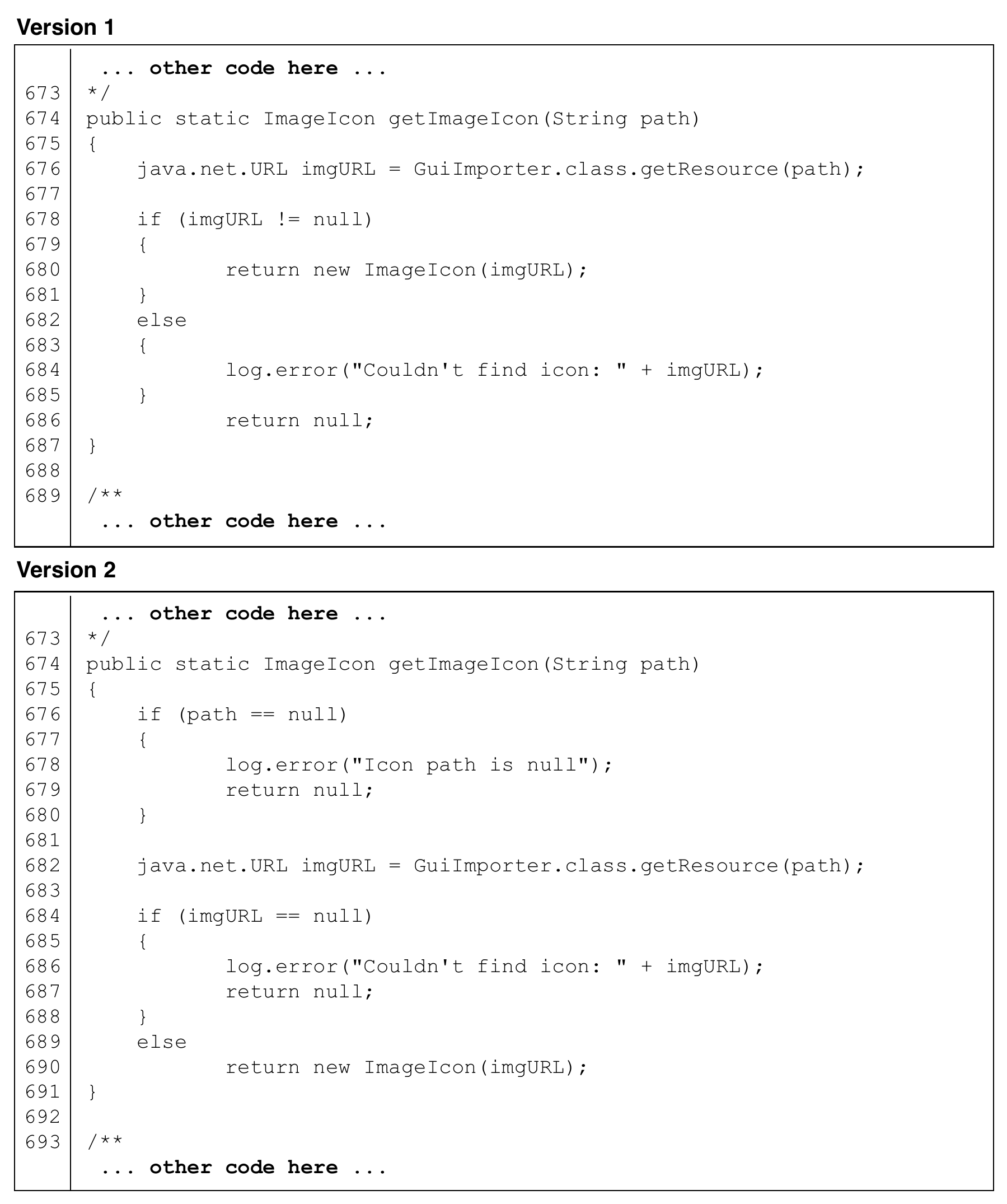}
    \caption{A set of changes from an older file into a newer file}
    \label{fig:changeofcode}
\end{figure}

\subsection{Myers}
\label{sec:myers}
\textit{Myers} algorithm was developed by \cite{Myers86ano(nd)}. 
In the \texttt{git diff} command, this algorithm is used as the default. 
The operation of this algorithm traces the two primary identical sequences recursively with the least edited script. 
Since the \textit{Myers} only notices the sequences which are actually equal in both, the comparison between the other prior and posterior subsequences is executed repetitively for the entire remaining sequences.

\begin{figure}
    \centering
    \subfloat[Step 1: pair up the first three matching lines (line 673-675 in both versions)\label{fig:myers1}]{
        \includegraphics[width=1\textwidth]{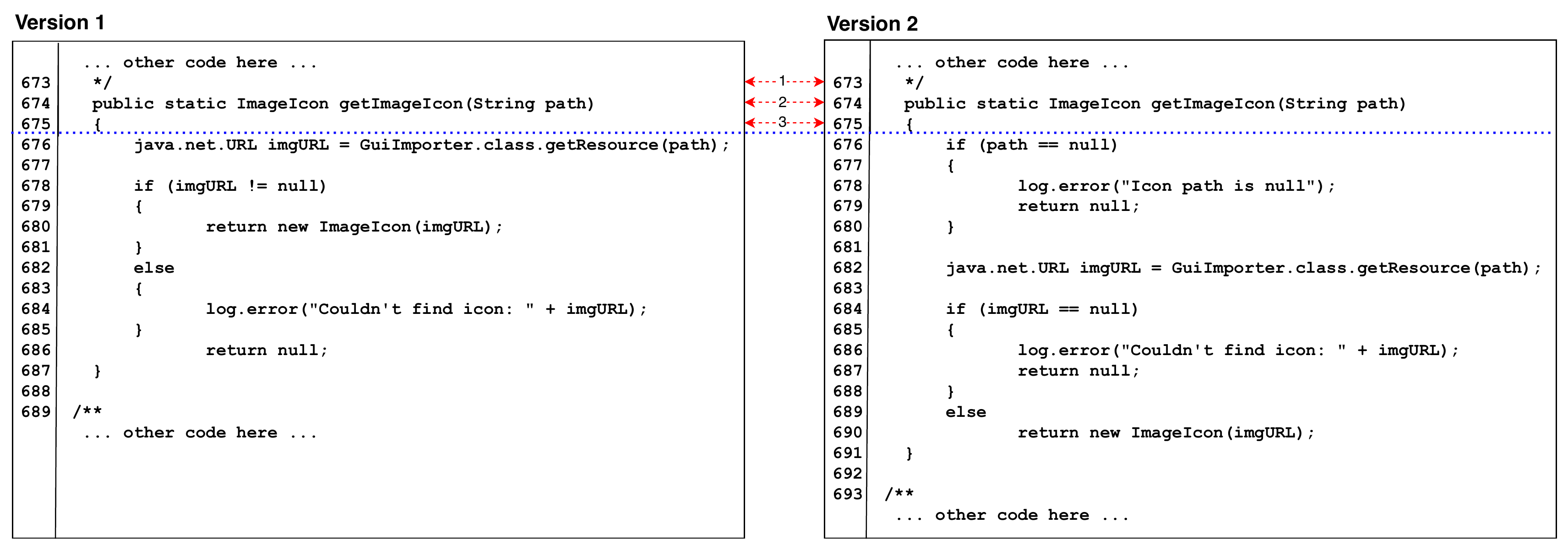}
    }
    \hfill
    \subfloat[Step 2: pair up the fourth matching lines (line 679 in version 1 and line 677 in version 2)\label{fig:myers2}]{
        \includegraphics[width=1\textwidth]{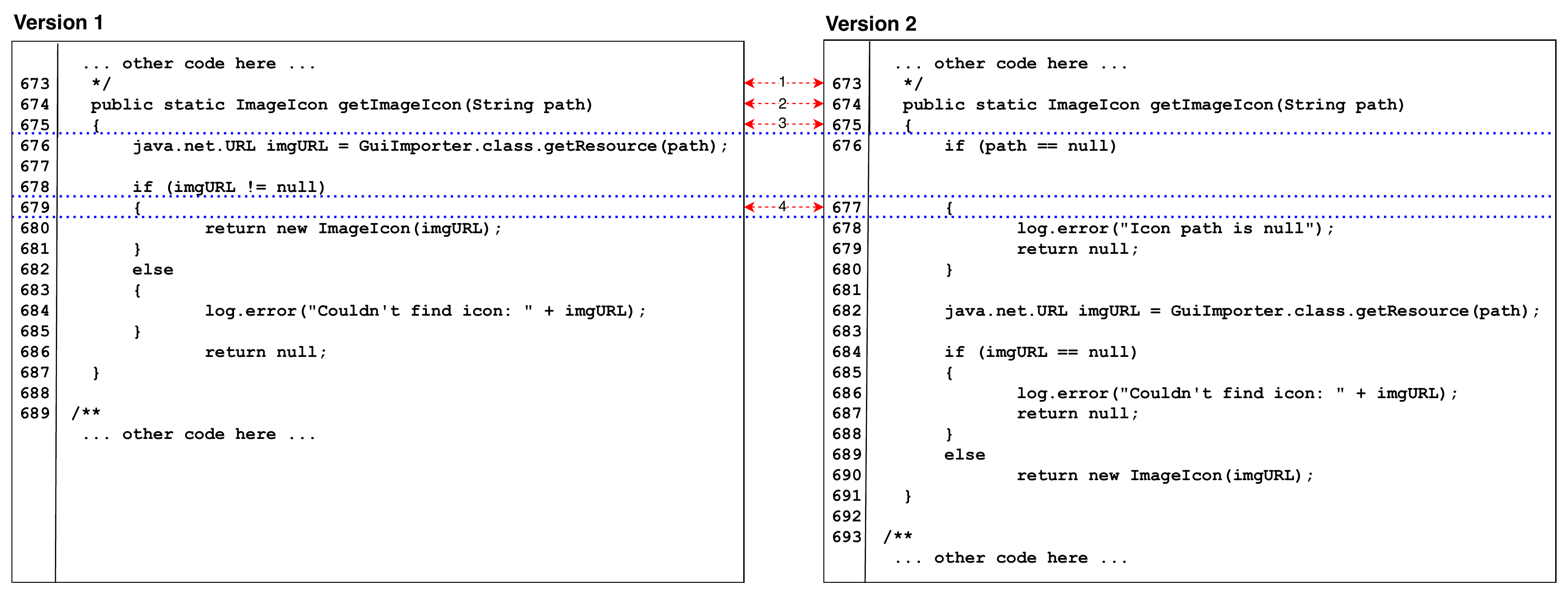}
    }
    \hfill
    \subfloat[Step n: final step after pairing up all matching lines\label{fig:myersn}]{
        \includegraphics[width=1\textwidth]{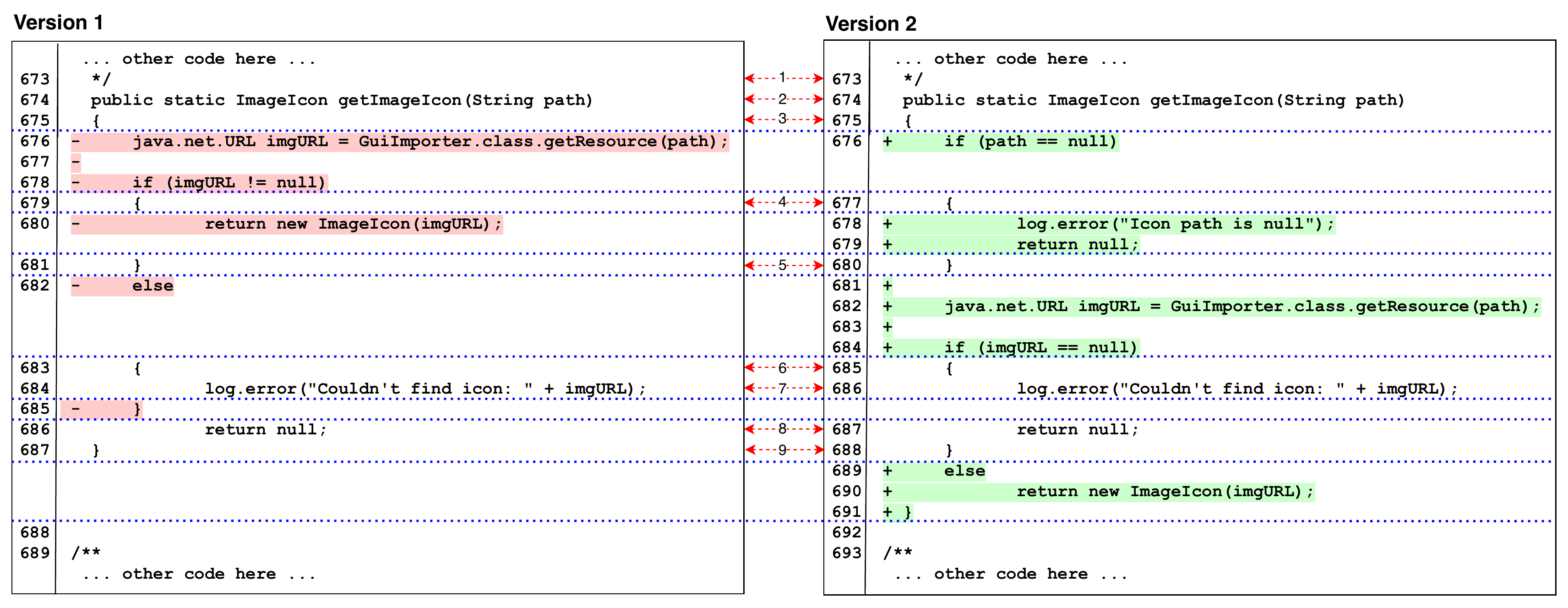}
    }
    \hfill
    \caption{How \textit{Myers} identifies the \texttt{diff}}
    \label{fig:myers_procedure}
\end{figure}

Figure \ref{fig:changeofcode} indicates several code changes from the first into the second version of the same file (\texttt{GuiCommonElements.java}) taken from Openmicroscopy project\footnote{\label{codeHbase}\url{https://github.com/openmicroscopy/openmicroscopy/commit/844e0fde447d2d069fb17c480e95acf4d372afc4#diff-07322c93ef4fb3f0dd245932b74b10e1}}.
As can be seen in the figure, the code between line 673 and 689 in the first version transformed to the newer version between line 673 and 693. 
Figure~\ref{fig:myers_procedure} shows how \textit{Myers} algorithm generates the \texttt{diff} output from the code changes in Figure~\ref{fig:changeofcode}.
First, the \textit{Myers} scans the lines of code sequentially from the first line in both versions of the same file to find a line pair that match up each other.
Once the exact same lines between the two versions of the file are found by the algorithm, the lines will be considered as the unmodified lines (e.g. pair of lines 673-675 in both versions in Figure~\ref{fig:myers1}).
The algorithm then do the same scanning to extract the other pairs of matched lines for the remaining lines of code repetitively, as depicted in Figure~\ref{fig:myers2} and Figure~\ref{fig:myersn}.
In Figure~\ref{fig:myersn}, we can see all unmodified lines found by the \textit{Myers} algorithm: pair of line 673-675 in both versions, pair of line 679 in Version 1 and 677 in Version 2, 681 and 680, 683 and 685, 684 and 686, 686 and 687, and 687 and 688).
The unpaired lines in Version 1 are subsequently considered as the deleted lines, while the unpaired lines in Version 2 are counted as the added lines.
As a result, the \textit{Myers} algorithm produces the paired and unpaired lines from the first and second version of the same file in sequence, as illustrated in Figure~\ref{fig:myersoutput}.

The \textit{Minimal} algorithm is the extended version of \textit{Myers}. 
The operation of this algorithm in finding the changes resulted from a comparison of two objects resembling the \textit{Myers}, but an extra attempt was made to keep the patch size as minimal as possible\footnote{\label{minimalweb}\url{http://fabiensanglard.net/git_code_review/diff.php}}. 
As a result, the \texttt{diff} lists created using this algorithm are often identical with the \textit{Myers}.
If we apply the \textit{Minimal} algorithm to the code in Figure \ref{fig:changeofcode}, the \texttt{diff} output is shown in Figure \ref{fig:myersoutput} as well.

A major limitation of the \textit{Myers} algorithm is it frequently catches the blank lines or parentheses and conforms the lines to match instead of catching the line that is ``unique'' (i.e. lines that occur exactly once or the least occurrence in both versions), such as code of function declaration, or a line of assignment.
Consequently, the \textit{Myers} sometimes produces unclear \texttt{diff} lists that do not describe the actual code changes. 
The position between changed code and code that replace them is often written distantly in inappropriate lines, or located separately in a line that does not represent the modification. 
Additionally, there is occasionally a conflict of identification of the changed code; for example, the code in lines 4 and 15 in Figure \ref{fig:myersoutput}. 
In fact, these lines of code were derived from the same unique line that was unmodified. 
Using the \textit{Myers} algorithm, this unique line is detected as a changed code even though it does not show the alteration. 
This makes it possible to cause misidentification of a code change.

\subsection{Histogram}
\label{sec:histogram}
The \textit{Histogram} algorithm is the enhanced version of \textit{Patience}, which
was built by Bram Cohen who is renowned as the BitTorrent developer\footnote{\label{patienceweb}\url{https://alfedenzo.livejournal.com/170301.html}}. 
It supports low-occurrence common elements which are applied to improve efficiency. 
The \textit{Histogram} was initially built in jgit\footnote{\label{jgit}\url{http://eclipse.org/jgit/}} and was introduced in \texttt{git 1.7.7}.

The \textit{Patience} marks the important lines within the text by focusing on the lines that have the smallest number of occurrences, but are essential. 
This \texttt{diff} automated procedure is an LCS-based problem as well, but it uses a different technique. 
The \textit{Patience} only notices the longest common subsequence of the marked lines attained from the lines which emerge uniquely in a specific range and the lines that are also written precisely similar in both files. 
This implies that the lines having a single bracket or a new line are usually disregarded; otherwise, the \textit{Patience} retains the distinctive line such as a function definition.

The \textit{Histogram} strategy works similarly to the \textit{Patience} by developing a histogram of the appearances for every line in the first version of a file. 
Every element in the second version is subsequently shown to match with the first sequence in an orderly way to find the existences of the elements and to count the occurrences. 
If the elements exist and their presences are less than in the first sequence, they are expected to be a potential LCS. 
Once the screening is finished for the second sequence, the lowest occurrence of LCS is marked as the separator. 
Two sections resulting from the partition (i.e. section 1 represents the area before the LCS, while section 2 represents the region after the LCS), are then executed repetitively using the same process as the beginning of the algorithm. 
This means that the \textit{Histogram} performs similarly to the \textit{Patience} if a unique common element exists in both files; otherwise, it selects the element that has the least occurrences. 
In comparison with the other two \texttt{diff} algorithms, (i.e. the \textit{Myers} and the \textit{Patience}), the \textit{Histogram} nevertheless, has been declared much quicker\footnote{\label{histtest}\url{https://marc.info/?l=git&m=133103975225142&w=2}}.

\begin{figure}
    \centering
    \subfloat[Step 1: pair up the first and second matching unique lines (line 674 in both versions and line 673 at the upper section of the partition)\label{fig:histogram1}]{
        \includegraphics[width=1\textwidth]{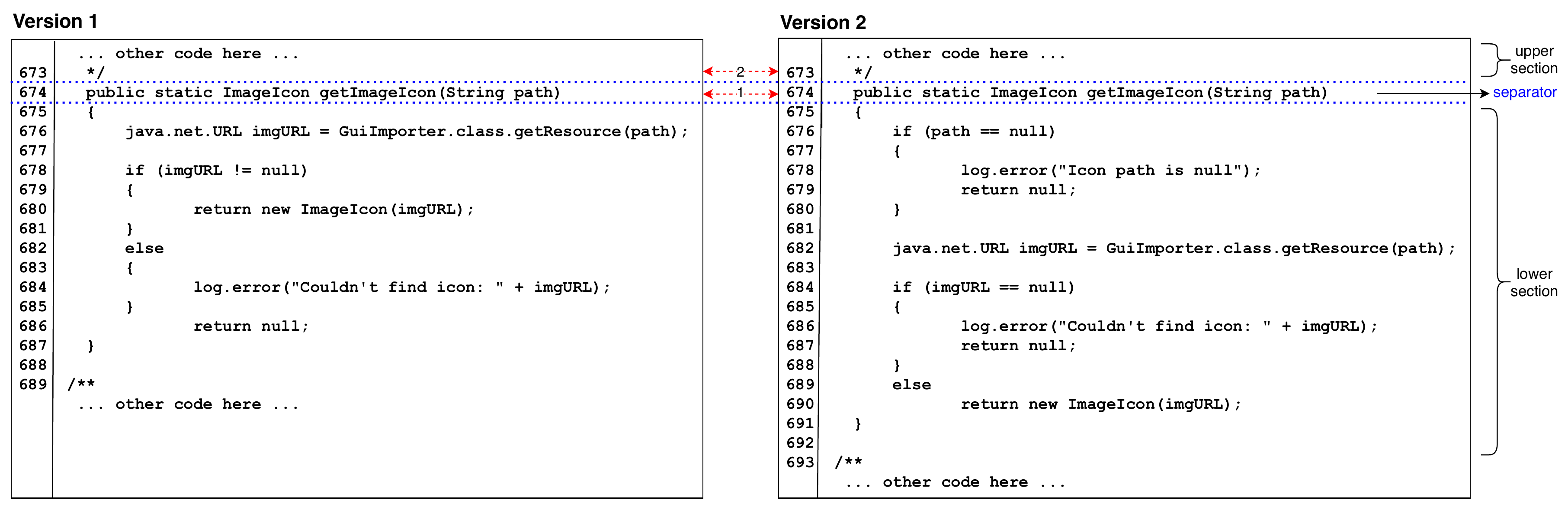}
    }
    \hfill
    \subfloat[Step 2: pair up the third matching unique line at the lower section of the partition (line 676 in version 1 and line 682 in version 2)\label{fig:histogram2}]{
        \includegraphics[width=1\textwidth]{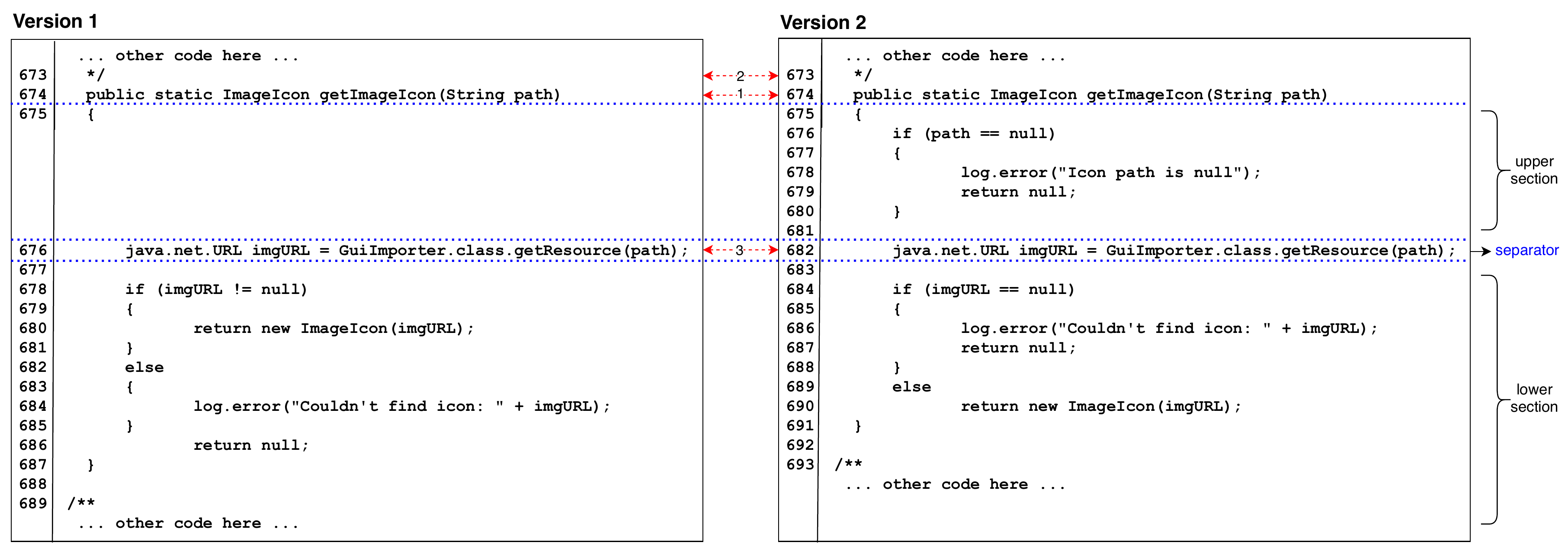}
    }
    \hfill
    \subfloat[Step n: final step after pairing up all matching unique lines\label{fig:histogramn}]{
        \includegraphics[width=1\textwidth]{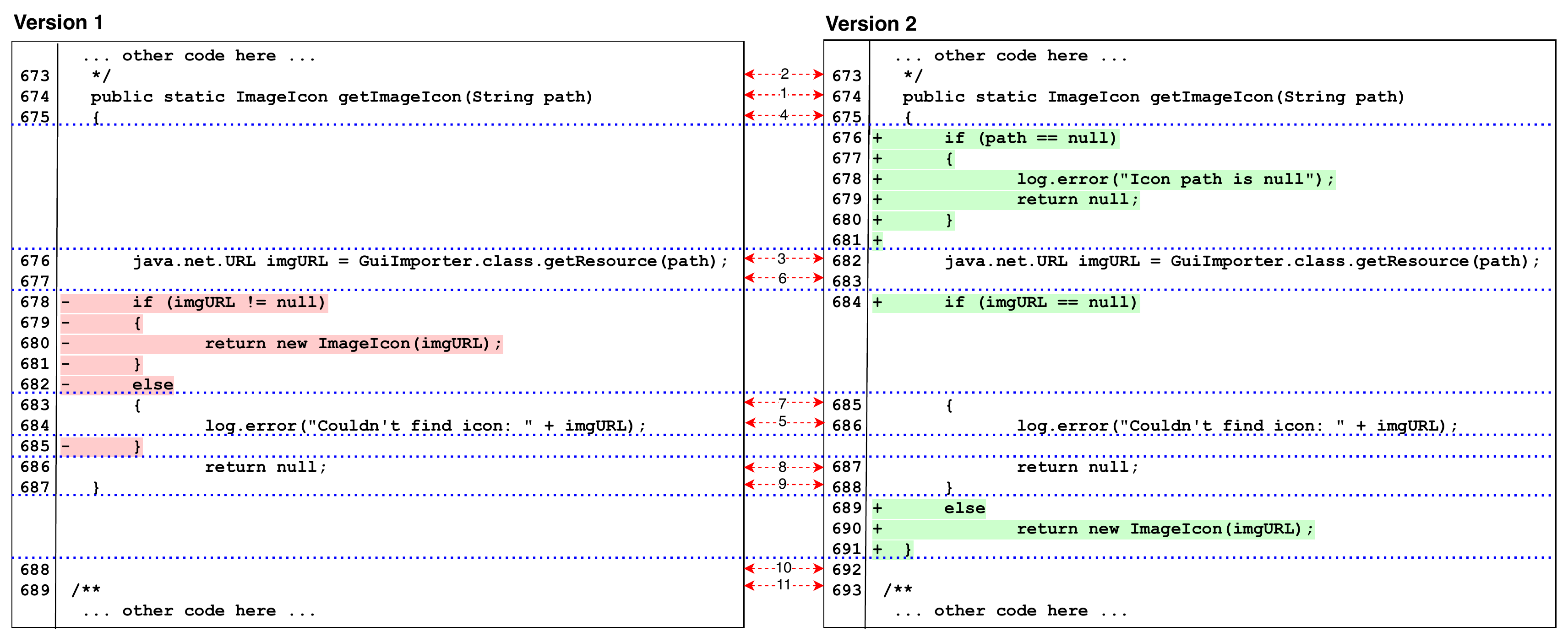}
    }
    \hfill
    \caption{How \textit{Histogram} identifies the \texttt{diff}}
    \label{fig:histogram_procedure}
\end{figure}


\begin{figure}
    \centering
    \subfloat[\textit{Myers'} diff\label{fig:myersoutput}]{
        \includegraphics[width=0.9\textwidth]{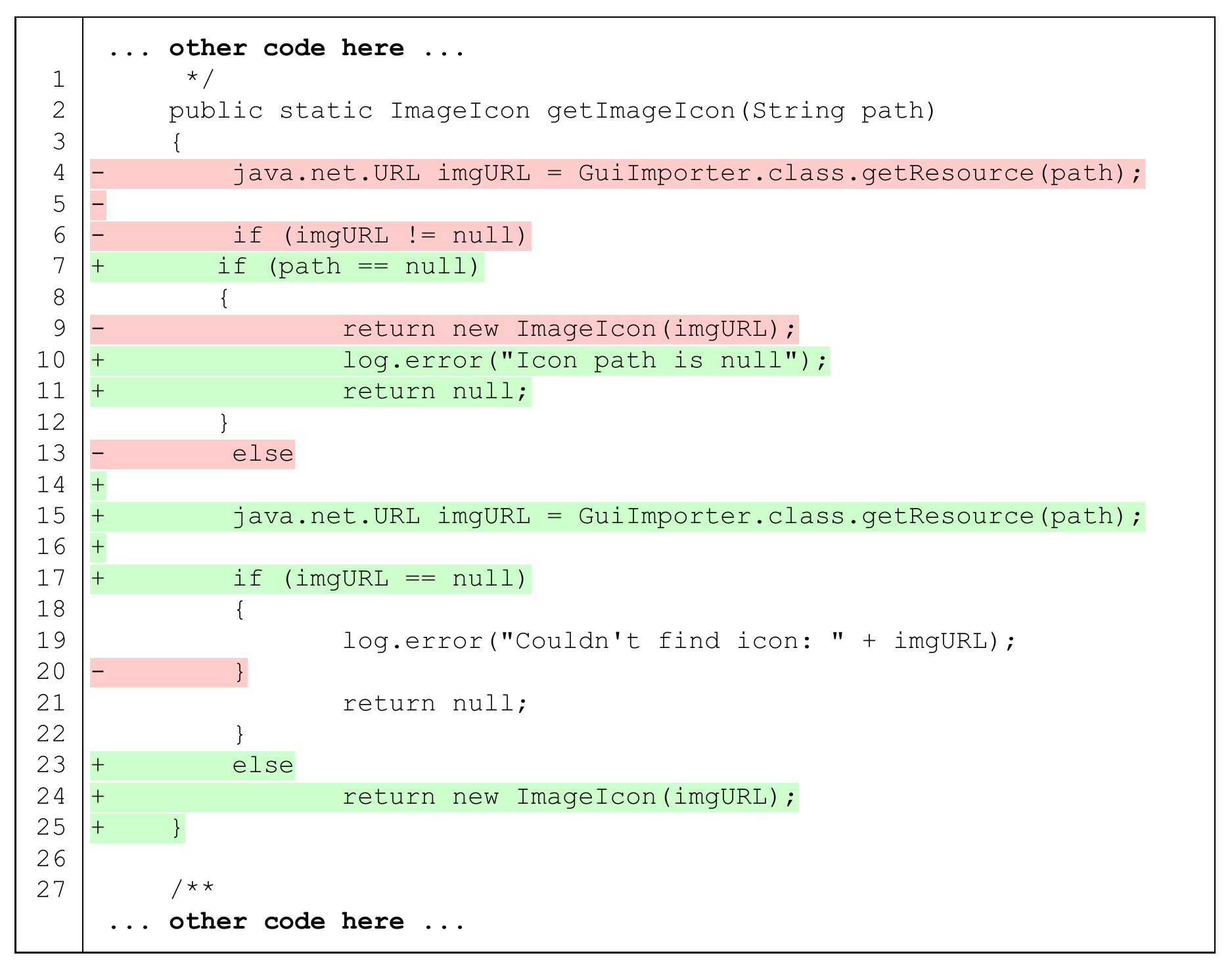}
    }
    \hfill
    \subfloat[\textit{Histogram'}s diff\label{fig:patienceoutput}]{
        \includegraphics[width=0.9\textwidth]{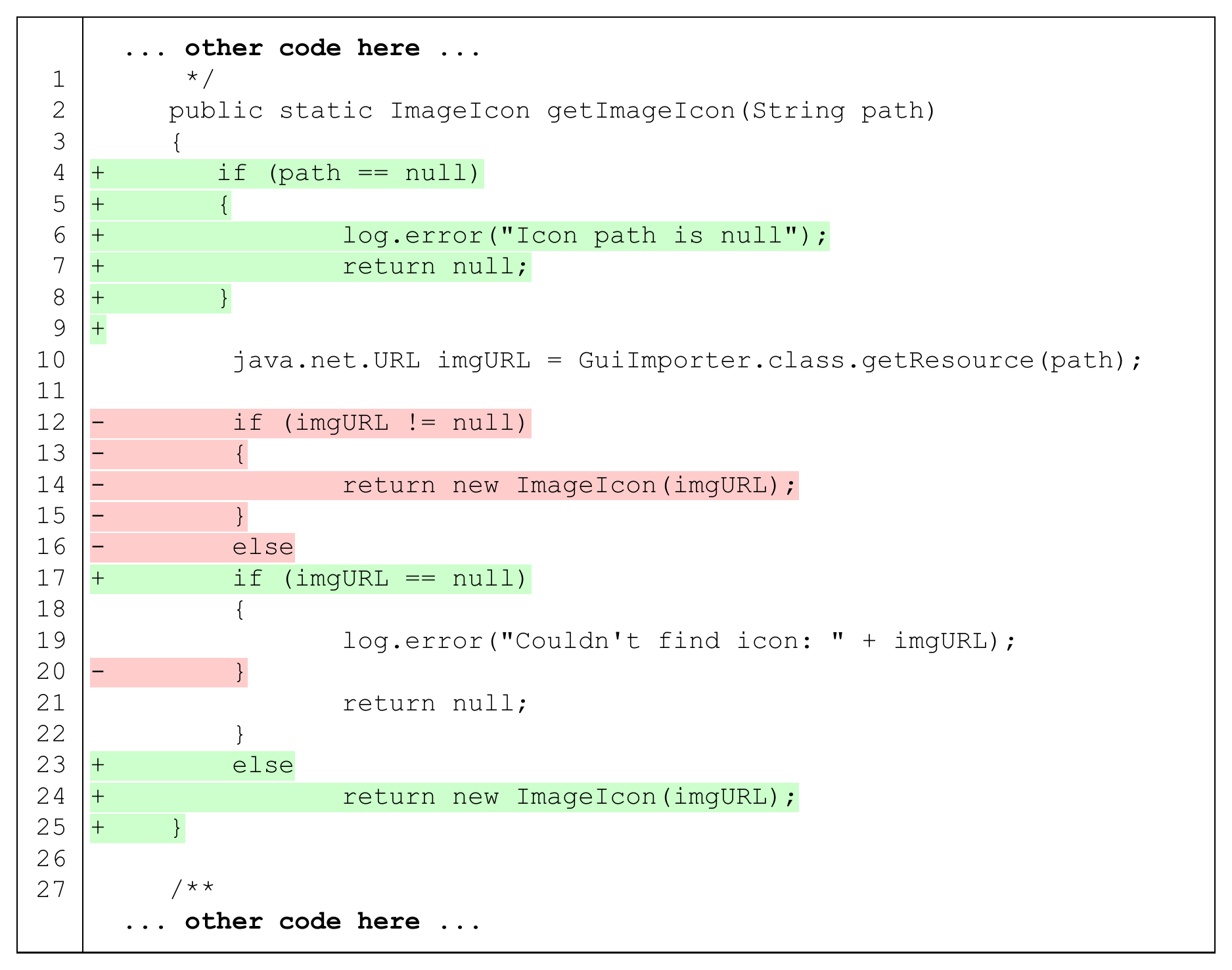}
    }
    \caption{\texttt{Diff} outputs produced by \textit{Myers} and \textit{Histogram}}
    \label{fig:outputs}
\end{figure}


To easily understanding the \textit{Histogram} generates the \texttt{diff} output from Figure~\ref{fig:changeofcode}, we describe the procedure in Figure~\ref{fig:histogram_procedure}.
First, the \textit{Histogram} scans all elements in the first version of the file to count the appearances of each line.
Every line in the second version is extracted to match with the element in the first version sequentially to find the exact same line and count the occurrences.
If the algorithm found the lines in both versions are match and their presences are unique (i.e. occurs exactly once or have the lowest occurrences in both), they are considered as the potential LCS which is then marked as the separator.
As shown in Figure~\ref{fig:histogram1}, line 674 in both versions are marked as the first separator.
Two sub-sections are created after this slicing, that is, the area before and after the separator.
Within those sub-sections, the algorithm find more unique pairings; lines that are not unique when scanning the entire document can be unique when the algorithm consider a sub-section.
The same process is then applied to both sub-sections.
The \textit{Histogram} compares line 673 in the upper section in both versions, and lines 675-689 in Version 1 with lines 675-693 in Version 2 in the lower sections.
Due to the least appearances of line 673 only in the upper section in both versions, thus, this line is expected to be the second separator.
In the lower section, the scanning process is re-executed from the beginning.
As illustrated in Figure~\ref{fig:histogram2}, the process yields a new separator (i.e. line 676 in Version 1 and line 682 in Version 2) and two new sub-sections (i.e. line 675 in Version 1 and line 675-681 in Version 2 as the upper section, and line 677-689 in Version 1 and line 683-693 in Version 2 as the lower section).
The same process is subsequently executed repetitively for the two new sub-sections resulting from the partition.
Figure~\ref{fig:histogramn} shows the final step after comparing all elements in both versions.
All potential LCS that are marked as the separator are expected to be the unmodified lines, while the other lines are considered as the deleted lines in Version 1 and the added lines in Version 2.
As a result, the \texttt{diff} output is generated as described in Figure~\ref{fig:patienceoutput}.

In contrast with the \textit{Myers}, the \textit{Histogram} algorithm provides \texttt{diff} results that are easier for software archives miners to understand, as the \textit{Histogram} more clearly separates the changed code lines. 
This algorithm splits the changed lines of code by trying to match up unique lines between two versions of the same file. 
Thus, it will reduce the occurrences of conflict (i.e. a line of an unchanged code identified as a changed code, so that in the \texttt{diff} list, this code is written in duplicate as both a deleted and inserted code). 
For example, if we extract the differences between the two versions of the same file in Figure \ref{fig:changeofcode} using the \textit{Histogram} in the \texttt{git diff} command, we obtain the output as depicted in Figure \ref{fig:patienceoutput}. 
A unique line of code in line 10 of Figure \ref{fig:patienceoutput} is not detected as a changed code due to its role as the benchmark to match the line, where this line is identified as a changed code in case of \textit{Myers}. 
This influences the sequences of the other changed code. 
An additional block of \textit{if} condition is written between lines 4 and 9 where it should be placed. 
This block of code is clearly understood as the new code inserted before the statement of the assignment code (code in line 10 which is used as one of some unique lines to match). 
It is also obvious that the code between lines 12 and 16 were replaced by one line of code in line 17, while the closing curly brace in line 20 was omitted from the files, and three new lines of code (line 23, 24 and 25) were added at the end of the code in Figure \ref{fig:patienceoutput}.

\section{Systematic Mapping: How Previous Studies Used Git Diff?}
\label{sec:survey}

To understand the ways in which the previous studies use \texttt{diff}, we conducted a systematic mapping of papers that used the \texttt{git diff} command for their studies. 
As described by \cite{Petersen:2008:SMS:2227115.2227123}, a systematic mapping study can provide and visualize a statistical insight of a study domain by classifying and quantifying the number of publications related to the research interest within the same study domain.
The main activity of the method was searching the relevant literature from a wide range of publications including journal articles, books, documented archives and scripts.

We performed a systematic mapping as we intend to: (i) draw an overview of the research area through quantification in a structured way~\citep{Kuhrmann:2017:PDL:3147777.3147805}, (ii) confirm the knowledge in the currently published studies~\citep{PETERSEN20151}. 
A systematic mapping is reliable because the findings are repeatable and consistent across the time~\citep{WOHLIN20132594}, and they are beneficial for better reporting of some empirical findings of the primary studies~\citep{Budgen_usingmapping}.

To understand how recent studies used \texttt{git diff}, we prepared the following research questions for this systematic mapping.
\begin{itemize}
    \item Which \textit{diff} algorithm is used?
    \item What kind of software artifact is analyzed, code or other documents?
    \item What are purposes of using \textit{diff}?
    \item Where does the data source come from, OSS or industry?
\end{itemize}

\subsection{Procedure}
\label{sec:procofsurvey}

\begin{figure}[ht]
    \begin{center}
        \includegraphics[width=1\textwidth]{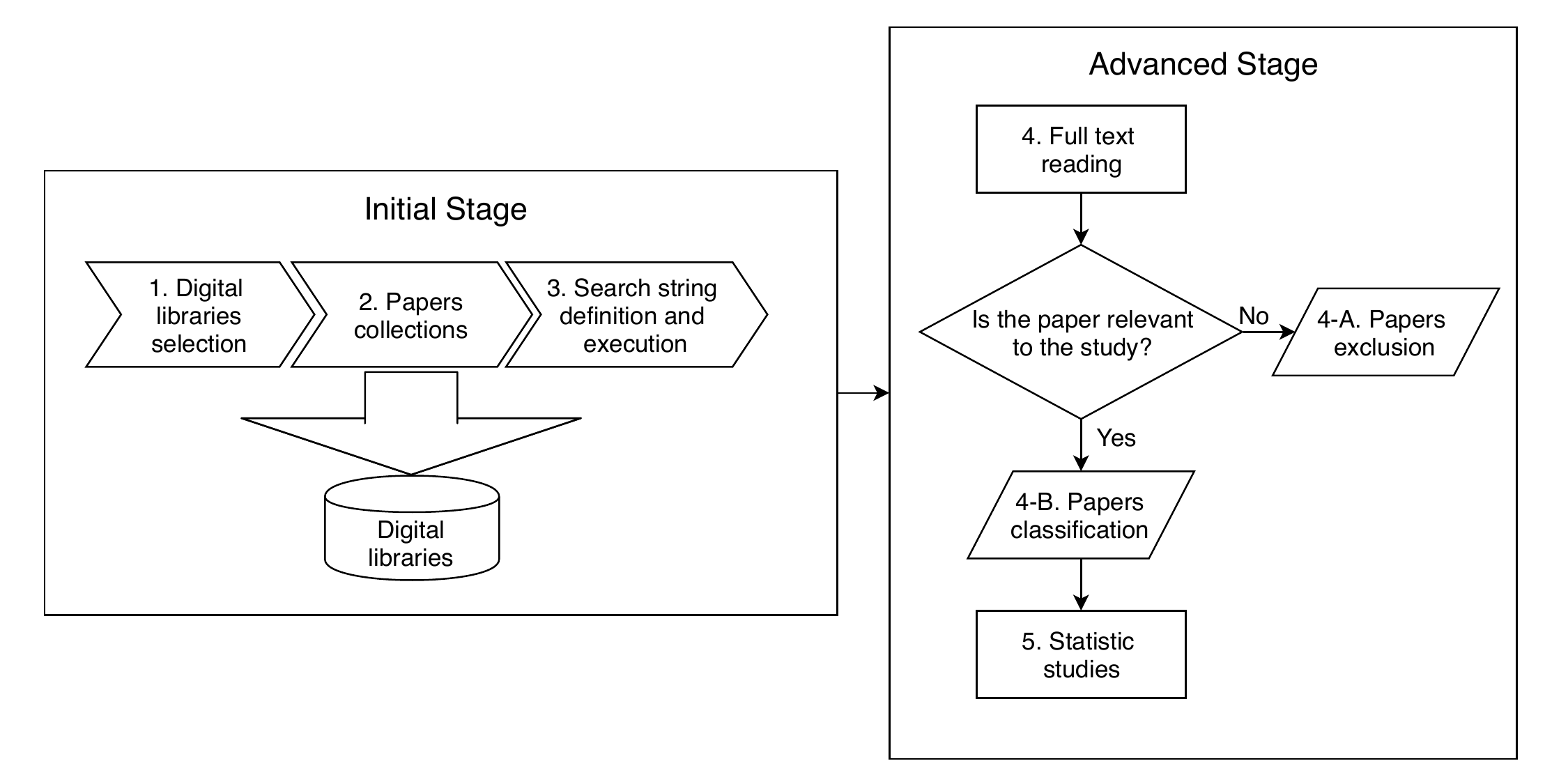}
    \end{center}
    \caption{Design of the Survey Procedure}
    \label{fig:surveyprocedure}
\end{figure}

Figure \ref{fig:surveyprocedure} illustrates an overview of our systematic mapping procedure, which is divided into an initial stage and an advanced stage. 
The first stage has three steps including a digital libraries selection, papers collection, search string definition and initial search execution. 
The second stage begins with repetitive manual exclusion by narrowing the search terms and the reading of full papers, followed by paper classification, and statistical analyses.

\textbf{Step 1: Digital Libraries Selection}.
The selection of appropriate literature is essential to guarantee high-quality papers and to grasp the state-of-the-art issues in the software engineering field~\citep{Kavitha_collectiondevelopment}. 
We specifically targeted papers which were published in high ranking journals and conference proceedings of the software engineering area. 
To maximize the probability of finding highly relevant good quality articles, we used three specific digital resources: ACM Digital Library\footnote{\label{ACM}\url{https://dl.acm.org/}}, IEEE Xplore\footnote{\label{ieee}\url{https://ieeexplore.ieee.org/}}, and SpringerLink\footnote{\label{springer}\url{https://link.springer.com/}}. 
Table \ref{tab:journalandconf} shows the list of the publication sources used in our survey including their impact factors (IF)\footnote{\label{impactfactor}\url{https://www.scimagojr.com/}} and rankings published in 2018 CORE Rankings\footnote{\label{coreranks}\url{http://www.core.edu.au/conference-portal/2018-conference-rankings-1}}. 
We gathered published papers from these three digital sources between the years of 2013 and 2017.

\begin{table}[]
    \caption{List of Surveyed SE Journals and Conferences}
    \label{tab:journalandconf}
    \begin{tabularx}{\linewidth}{lp{7.8cm}l}
        \hline\noalign{\smallskip}
        Category & Name of Journal or Conference & IF or Rank	\\
        \noalign{\smallskip}\hline\noalign{\smallskip}
        Journal & IEEE Transactions on Software Engineering (TSE) & IF = 3.331	\\
        & Empirical Software Engineering (EMSE) & IF = 2.933	\\
        & ACM Transactions on Software Engineering and Methodology (TOSEM) & IF = 1.946 \\
        Conference & ACM Conference on Object Oriented Programming Systems Languages and Applications (OOPSLA) & Rank = A* \\
        & ACM-SIGPLAN Conference on Programming Language Design and Implementation (PLDI) & Rank = A*   \\ 
        & International Conference on Software Engineering (ICSE) & Rank = A*	\\
        & Joint Meeting of the European Software Engineering Conference and the ACM SIGSOFT Symposium on the Foundations of Software Engineering (ESEC/FSE) & Rank = A*	\\
        & Automated Software Engineering (ASE) & Rank = A \\
        & International Conference on Software Maintenance and Evolution (ICSME) & Rank = A	\\
        & International Conference on Mining Software Repositories (MSR) & Rank = A	\\
        & International Symposium on Software Testing and Analysis (ISSTA) & Rank = A  \\
        \hline
    \end{tabularx}
\end{table}

\begin{figure}
    \begin{center}
        \includegraphics[width=1\textwidth]{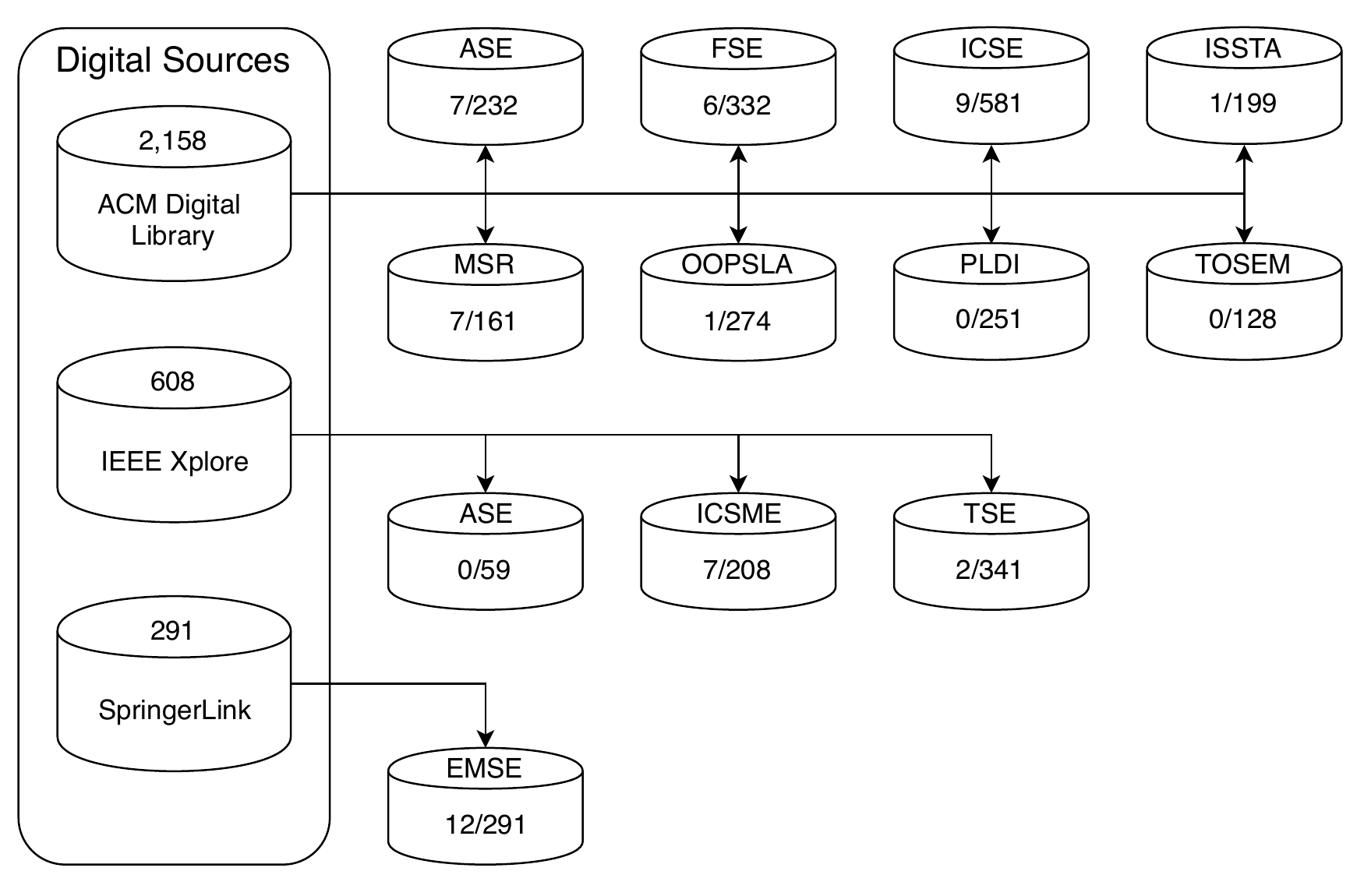}
    \end{center}
    \caption{Number of collected papers from each source}
    \label{fig:papercollection}
\end{figure}

\textbf{Step 2: Papers Collection}.
To reduce bias in the context of the study, we only collected technical papers. 
Papers which did not meet our criteria (i.e shorter-than-10-page papers, editorials, panels, poster sessions, and opinions) were excluded. 
As depicted in Figure \ref{fig:papercollection}, by applying our criteria, we sourced 3,057 papers in total from the three digital sources in a 5-year time span. 

\textbf{Step 3: Search String Definition and Execution}.
In this step, we formulated search keywords to filter the targeted papers into more specific works that use the \texttt{git diff} command. 
We defined three specific search terms related to the command, namely \textit{git}, \textit{log} and \textit{diff}. 
Papers that contained one of three words with an exact match without affixes or suffixes (e.g. \textit{github, blog, logarithm, logging, different, difficult} etc.) were collected. 
Since we only focus on the study that used \texttt{diff} command in \texttt{git} repositories, papers that do not exactly mention at least one of the three keywords are excluded despite they use other terms such as \textit{differencing} which might indicates the implementation of the other \texttt{diff} tools.
The command \texttt{git log} was also targeted because this command can produce \textit{diff} with specific options.
By using these three search terms, all papers extracted from the databases were then manually scanned in full text. 
Consequently, only published works containing these three search strings were included. As a result of Step 3, we were able to identify 137 papers.

\textbf{Step 4: Full Text Reading}.
To ensure the collected previous studies are relevant to our objectives, we then performed a full text reading of the papers. 
This process was undertaken by the first and the second authors to avoid obscurity and to separate the primary studies more exhaustively based on their contents.
We applied the inclusive and exclusive criteria to the full paper which is described in Table \ref{tab:includexcludcriteria}. 
Papers that fit the inclusive criteria were kept for further processing while other papers that met the exclusive criteria were excluded from the study. 
After this step, we had 52 papers.

\begin{table}[]
    \caption{Inclusive and Exclusive Criteria}
    \label{tab:includexcludcriteria}
    \begin{tabularx}{\linewidth}{lp{9.8cm}}
        \hline\noalign{\smallskip}
        Group & Criteria	\\
        \noalign{\smallskip}\hline\noalign{\smallskip}
        Inclusive & Paper mentions the \texttt{git} command.  \\
        & Paper applies the \texttt{git} command to extract the data from \texttt{Git} repositories for further analyses in support of their works. \\
        Exclusive & Paper does not use the \texttt{git} command as part of their studies. For example, \texttt{git diff} is used only for motivating examples.	\\
        \noalign{\smallskip}\hline
    \end{tabularx}
\end{table}

\subsection{Results of the Mapping}
\label{sec:resultofsurvey}

\begin{figure}[]
    \center
    \includegraphics[width=1\textwidth]{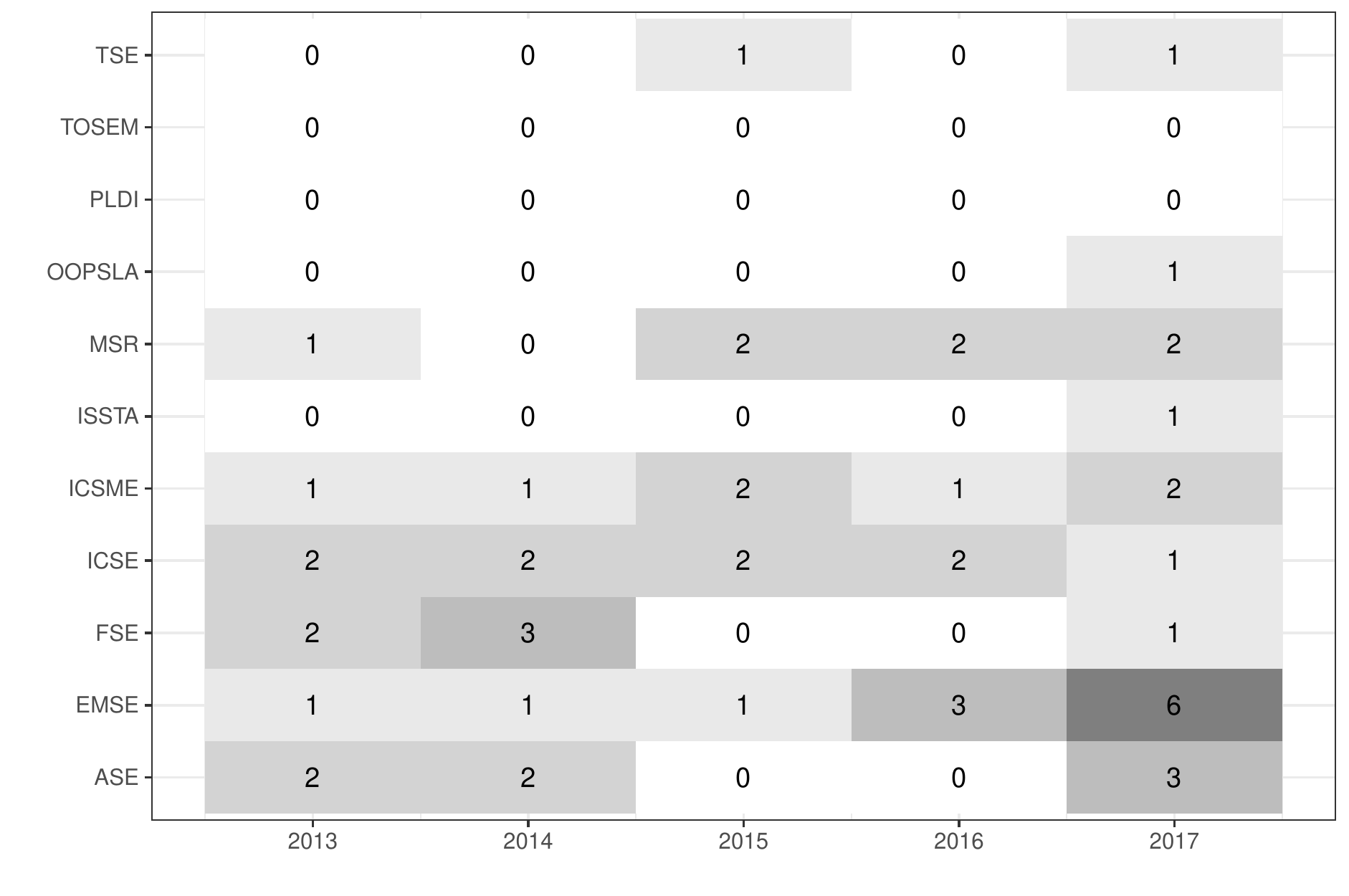}
    \caption{Number of papers per journals and conferences between 2013 and 2017}
    \label{fig:heatmap}
\end{figure}

Figure \ref{fig:heatmap} indicates the distribution of the number of papers in each journal and conference in the last 5 years. 
As can be seen in the heat map, all journals and conference proceedings published the works related to the \texttt{git diff} command application in at least one paper in 5 years except for the PLDI and TOSEM.  
Most papers that applied the \texttt{git diff} command are published on EMSE especially in 2017, accounting for 6 papers.

\subsubsection{Which \texttt{Diff} Algorithm Is Used?}
\label{sec:RQ1result}

Out of the 52 primary studies,we identified the application of different \texttt{diff} algorithms in the command in extracting the changes. 
Of particular note is that even though most instructions applied different options in the use of the \texttt{git} command to extract the required data, none of the previous selected works considered different \texttt{diff} algorithms.
This shows that all of the collected studies used \textit{Myers} as the default algorithm.

\subsubsection{What Kind of Software Artifact Is Analyzed?}
\label{sec:RQ2resultofparameter}

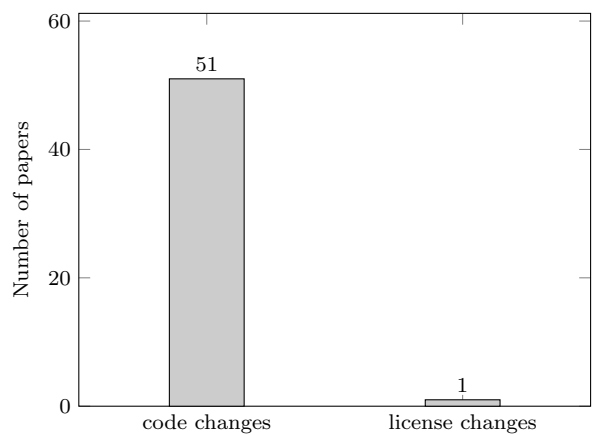
\begin{figure}[]
    \center
    \begin{tikzpicture}
        \begin{axis}[
        	compat=newest,
        	symbolic x coords={code changes, license changes},
            width=0.7\textwidth,
            height=0.35\textheight,
            xtick=data,
            ylabel={Number of papers},
            ymin = 0,
            bar width=28pt,
            nodes near coords,
            enlarge y limits={upper,value=0.2},
            enlarge x limits=0.5
            ]
            \addplot[ybar, fill=black!20] coordinates {
                            (code changes,51)
                            (license changes,1)
                        };
        \end{axis}
    \end{tikzpicture}
    \caption{Number of papers based on parameter searched using \texttt{git} command}
    \label{fig:searchparameter}
\end{figure}

To understand the components that were extracted using the \texttt{git} command in the previous studies, two main focuses emerged as our parameters to classify the documents; namely, code changes and license changes as depicted in Figure \ref{fig:searchparameter}. 
As can be seen in the figure, code changes were prominently the focus for researchers in extracting software repositories using the \texttt{git} command over five years. 
Thus, in our comparisons we analyze code changes extracted from the data source.

\subsubsection{What Are Purposes of Using \texttt{Diff}?}
\label{sec:RQ3resultofutility}

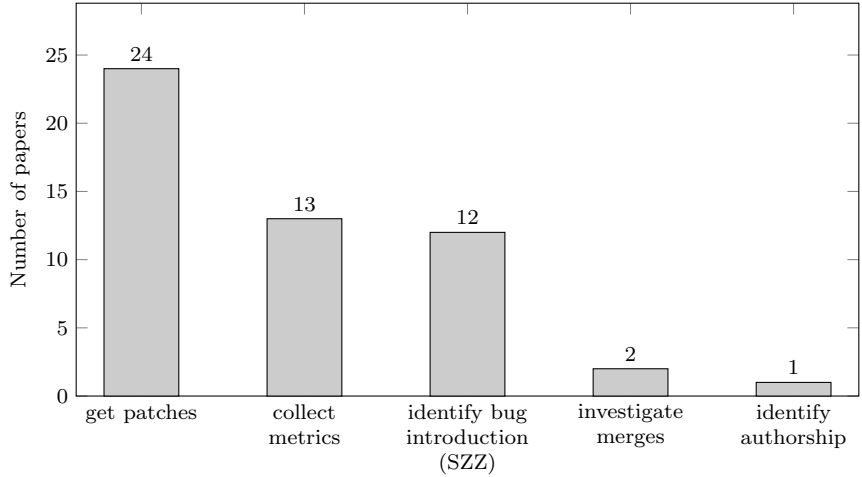
\begin{figure}[ht]
    \center
    \begin{tikzpicture}
        \begin{axis}[
            	compat=newest,
            	symbolic x coords={get patches, collect metrics, identify bug introduction (SZZ), investigate merges, identify authorship},
                width=1\textwidth,
                height=0.35\textheight,
                xtick=data,
                ylabel={Number of papers},
                ymin = 0,
                bar width=28pt,
                nodes near coords,
                enlarge y limits={upper,value=0.2},
                enlarge x limits=0.1,
                x tick label style={align=center, text width=1.7cm, rotate=0}
            ]
            \addplot[ybar, fill=black!20] coordinates {
                            (get patches,24)
                            (collect metrics,13)
                            (identify bug introduction (SZZ),12)
                            (investigate merges,2)
                            (identify authorship,1)
                        };
        \end{axis}
    \end{tikzpicture}
    \caption{Number of papers classified with the purpose of using the \texttt{git} command}
    \label{fig:utilityofgitcommand}
\end{figure}

By reading the papers manually, we summarized the purposes from the extraction of software development records and grouped them into five categories, as can be seen in Figure \ref{fig:utilityofgitcommand}. 

From the figure, we see that the most common purposes is to \textit{get patches}, amounting to as many as 24 studies, followed by \textit{collecting metrics} and \textit{identifying bug-introductions}, which covered 13 and 12 studies, respectively. 
A few studies addressed \textit{merges investigation} and \textit{authorship identification}. 
This finding motivated us to carry out a further investigation of the impact of different \texttt{diff} algorithms in the extraction of the added and deleted lines for metrics collection, bug-introducing change identification, and getting the patches.

\subsubsection{Where Does the Data Source Come From?}
\label{sec:RQ4resultofdataset}

\begin{figure}[]
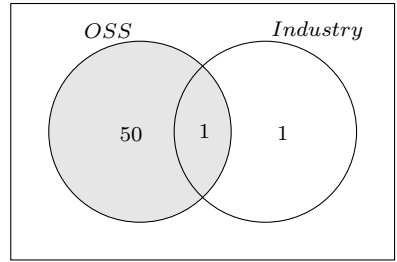

    \center
    \begin{venndiagram2sets}[labelA=, labelB=, shade=black!10]
        \fillA
        \setpostvennhook
        {
        	\draw (labelA) ++(150:4ex) node{$ OSS $};
            \draw (labelA) ++(-95:11ex) node[above]{$ 50 $};
            \draw (labelB) ++(20:6ex) node{$ Industry $};
            \draw (labelB) ++(-80:11ex) node[above]{$ 1 $};
            \draw (labelB) ++(-122:12.5ex) node[above]{$ 1 $};;
        }
    \end{venndiagram2sets}
    \caption{Distribution of the type of data sources used in prior studies}
    \label{fig:typeofdatasource}
\end{figure}

Our intention is to provide a comprehensive understanding of the different outcomes generated by different \texttt{diff} algorithms; thus, we need to run a set of tests of the algorithms' implementations in the \texttt{git diff} command. 
From the result of our dataset classification, open source software (OSS) is found to be dominated as the data source over the industrial type as illustrated in Figure \ref{fig:typeofdatasource}. 
Therefore, we mine the data from OSS projects to support our comparisons.

\subsection{Summary}

The survey results of the usage of the \texttt{git diff} command confirm that the previous studies conducted between 2013 and 2017 did not use various \texttt{diff} algorithms to extract the differences between the first and the second versions of the same file. 
In mining the \texttt{diff} lists, they applied the standard commands using a default \texttt{diff} algorithm with some additional options, but without considering various \texttt{diff} algorithms. 
We also found that the information most sought after in prior studies was code changes in open source projects. 
The code changes were mostly utilized to thoroughly investigate counting the number of line changes and to record them in the form of metrics, locating the origin of a bug using a specific method (i.e. SZZ algorithm), and analyzing the patches. 
The results of these types of analyses obviously rely on the \texttt{diff} records produced by an applied \texttt{diff} algorithm in the \texttt{git diff} commands.
Thus, different \texttt{diff} algorithms in extracting the line of code changes might differentiate the final result of a study and the conclusion of the description as well.

\section{Overview of Comparisons and Research Questions}
\label{sec:overviewandRQ}

The findings from our systematic mapping revealed the three most common purposes for using the \texttt{git diff} command. 
This encouraged us to undertake comparison analyses between the \textit{Myers} and \textit{Histogram} algorithms in three applications: metrics, the SZZ algorithm, and patches.
Our intention is to investigate the level of differences between the two \texttt{diff} algorithms used in these three applications and their possibility of affecting the result of studies. 
To achieve these goals, we address the following research questions:

\begin{itemize}[label={},itemindent=-2em,leftmargin=2em]
    \item \RqOne \\
    For metrics (Section~\ref{sec:metric}), equal and unequal changed lines in the files identified by the two \texttt{diff} algorithms were calculated based on two factors: the quantity and the position of the line of code.
    We then compared the quantity of the files that have the same and different added and deleted lines of code to understand the significance of the differences of both algorithms in providing the \texttt{diff} records.
    
    \item \RqTwo \\
    The result of locating bug-introducing changes using the SZZ algorithms relies on the \texttt{diff} results.
    In Section~\ref{sec:szz}, we applied the \textit{Myers} and \textit{Histogram} algorithms in the \texttt{git diff} command to know whether the \texttt{diff} lists affect the result of bug-introducing change identification.
    
    \item \RqThree \\
    Lastly, we compared the quality of the identified patches manually. 
    In Section~\ref{sec:patch}, we investigate 377 changes, a statistically representative sample of the 21,590 changes identified in the above two comparisons. 
\end{itemize}

In our three comparisons, to extract the changes, we apply the \texttt{git} command: \texttt{git diff -w --ignore-blank-lines --diff-algorithm=<algorithm> <parentcommit ID> <commit ID> -- <filename>.}
We use the same options \texttt{-w} and \texttt{--ignore-blank-lines} to ignore whitespace and the changes whose lines are all blank.
The use of various options is common according to the purposes to what extent the \texttt{diff} command generates the code changes.
However, since our focus is comparing \textit{Myers} and \textit{Histogram} as the \texttt{diff} algorithm that can be used at the same circumstances, we do not consider to investigate the impact of other options.

\section{Comparison: Metrics (RQ$_1$)}
\label{sec:metric}

\RqOne

\subsection{Analysis Design}
\label{sec:analysisdesign_metrics}

\begin{figure}[]
    \center
    \includegraphics[width=1\textwidth]{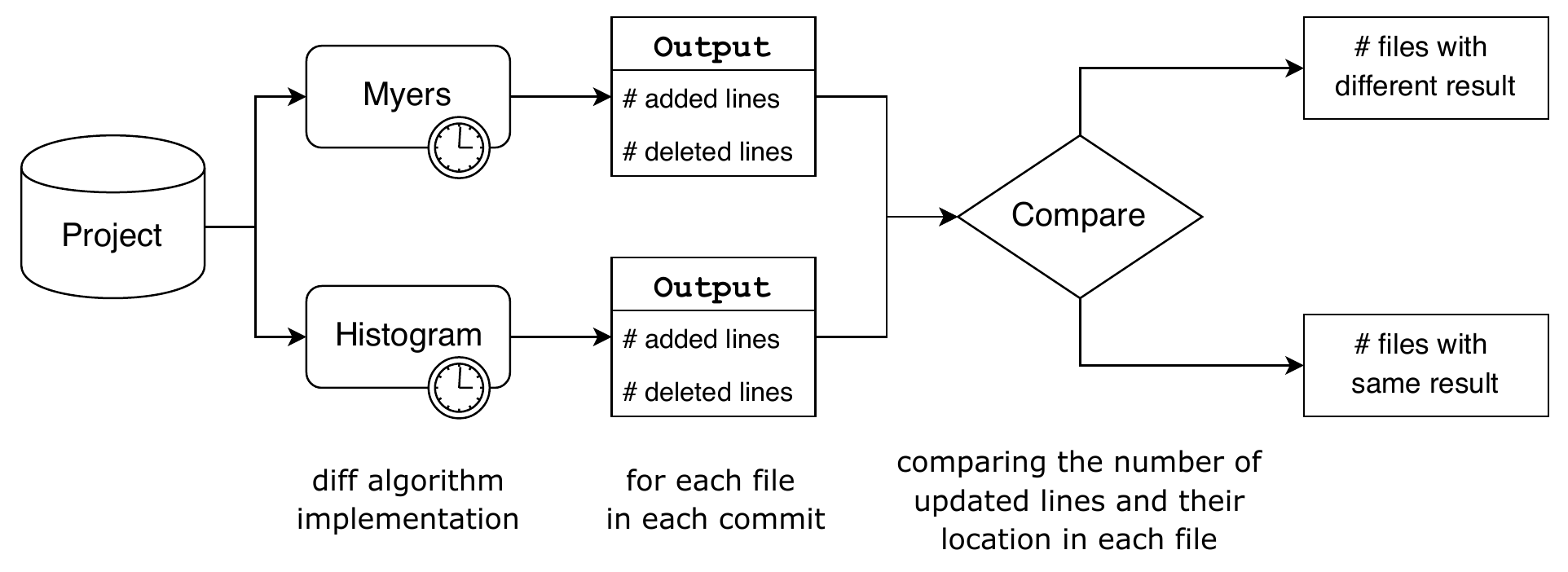}
    \caption{Overview of the metrics collection procedure}
    \label{fig:metricsteps}
\end{figure}

As illustrated in Figure \ref{fig:metricsteps}, we investigate the following two basic \texttt{diff}-related metrics with two \texttt{diff} algorithms: \textit{Myers} and \textit{Histogram}.
\begin{itemize}[label={}]
    \item \textbf{NLA} The number of added lines in a file.
    \item \textbf{NLD} The number of deleted lines in a file.
\end{itemize}

For our empirical analysis, we collected the Git repositories of 14 projects used in the previous study~\citep{Rausch:2017:EAB:3104188.3104231}, which are identified in our systematic mapping as a study utilizing \texttt{git} for collecting metrics. 
The targeted 14 projects are OSS that employ Continuous Integration (CI) and are written in Java. 
The descriptions of the projects and the number of commits in the \textit{master} branches are shown in Table~\ref{tab:CI-OSSProject}.

\begin{table}[]
    \caption{Targeted 14 open-source Java projects following the previous study~\citep{Rausch:2017:EAB:3104188.3104231}}
    \label{tab:CI-OSSProject}
    \begin{tabularx}{\linewidth}{lp{6.7cm}r}
        \hline\noalign{\smallskip}
        Project Name & Description & \#Commits	\\
        \noalign{\smallskip}\hline\noalign{\smallskip}
        Apache Storm & Distributed realtime computation system & 9,317	\\
        Butterknife & View binding library for Android & 836	\\
        Crate & Distributed SQL database & 8,646	\\
        Hystrix & Interactions controller library between distributed systems & 2,106	\\
        JabRef & Reference manager application that uses BibTex & 11,940	\\
        jcabi-github & Object oriented wrapper of GitHub API & 2,521	\\
        Openmicroscopy & Application to store biological microscopy light data in a standard format & 46,543	\\
        Presto & Distributed SQL query engine for big data & 13,561	\\
        RxAndroid & RxJava bindings for Android & 461	\\
        SpongeAPI & A minecraft plugin API & 2,479	\\
        Spring Boot & A framework to create Java applications & 17,087	\\
        Square OkHttp & An HTTP+HTTP/2 client for Android and Java applications & 3,171	\\
        Square Retrofit & An open source library to make HTTP communication simpler & 1,576 	\\
        WordPress-Android & WordPress for Android OS & 30,295	\\
        \hline
    \end{tabularx}
\end{table}

We investigated all modified files in all commits in the master branches.
To extract the NLA and NLD from the file, we implement the \texttt{git} command: \texttt{git diff -w --ignore-blank-lines --diff-algorithm=<algorithm> <parent commit ID> <commit ID> -- <filename>}.
We considered the results the same if the values of both NLA and NLD were the same with the two algorithms; otherwise, the results were considered different. 
However, several software engineering tasks that rely on such metrics do not consider the position of the added and deleted lines, where different position of the changed lines can be occurred by chance despite the same metrics value.
We conjecture that different number and position of changed lines can have different impact on empirical studies. 
Thus, we investigated the disagreement of the identified change locations separately. 
If the positions of each changed line of code were the same, we considered the results the same; otherwise, the results were considered different. 
File-level and commit-level results are discussed to see how the different results can appear in a different granularity.

\subsection{Results}
\label{sec:result_metrics}
Table \ref{tab:numberoffilebasedonLOC} summarizes the result from the comparison between two \texttt{diff} algorithms in 14 projects. 
From the total number of modified files identified by both algorithms, we counted the quantity of files in each commit that have same or different number values of NLA and NLD metrics. 
Similarity, the number of same and different results in changed locations are shown in the table.

\begin{table*}[ht]
    \caption{Total number of files that have the same and different values in metrics (NLA and NLD) and the position of changes.}
    \label{tab:numberoffilebasedonLOC}
    \centering
    \begin{threeparttable}
        \resizebox{\columnwidth}{!}{%
        \begin{tabular}{lrrrrrrrrrr}
            \hline\noalign{\smallskip}
            \multirow{2}{*}{Project} & \multirow{2}{*}{\#Files} & \multicolumn{4}{c}{Metrics (NLA and NLD)} & \multicolumn{4}{c}{Locations of Changes}	\\
            \cline{3-10}
            \noalign{\smallskip}
            & & \multicolumn{2}{c}{\#Same\hspace{8mm} \%} & \multicolumn{2}{c}{\#Different \%} & \multicolumn{2}{c}{\#Same\hspace{8mm} \%} & \multicolumn{2}{c}{\#Different \%} \\
            \noalign{\smallskip}\hline\noalign{\smallskip}
            Apache Storm & 22,011 & 21,278 & 96.7\% & 733 & 3.3\% & 20,979 & 95.3\% & 1,032 & 4.7\% \\
            Butterknife & 1,873 & 1,804 & 96.3\% & 69 & 3.7\% & 1,774 & 94.7\% & 99 & 5.3\%	\\
            Crate & 44,463 & 43,522 & 97.9\% & 941 & 2.1\% & 42,723 & 96.1\% & 1,740 & 3.9\%	\\
            Hystrix & 3,310 & 3,192 & 96.4\% & 118 & 3.6\% & 3,097 & 93.6\% & 213 & 6.4\%	\\
            JabRef & 55,988 & 54,375 & 97.1\% & 1,613 & 2.9\% & 53,609 & 95.7\% & 2,379 & 4.3\%	\\
            jcabi-github & 6,218 & 6,170 & 99.2\% & 48 & 0.8\% & 6,131 & 98.6\% & 87 & 1.4\%	\\
            Openmicroscopy & 118,349 & 115,126 & 97.3\% & 3,223 & 2.7\% & 112,548 & 95.1\% & 5,801 & 4.9\%	\\
            Presto & 73,572 & 72,471 & 98.5\% & 1,101 & 1.5\% & 71,455 & 97.1\% & 2,117 & 2.9\%	\\
            RxAndroid & 627 & 613 & 97.8\% & 14 & 2.2\% & 603 & 96.2\% & 24 & 3.8\%	\\
            SpongeAPI & 10,757 & 10,584 & 98.4\% & 173 & 1.6\% & 10,395 & 96.6\% & 362 & 3.4\%	\\
            Spring Boot & 62,137 & 60,805 & 97.9\% & 1,332 & 2.1\% & 60,214 & 96.9\% & 1,923 & 3.1\%	\\
            Square OkHttp & 7,345 & 7,206 & 98.1\% & 139 & 1.9\% & 7,108 & 96.8\% & 237 & 3.2\%	\\
            Square Retrofit & 3,473 & 3,394 & 97.7\% & 79 & 2.3\% & 3,351 & 96.5\% & 122 & 3.5\%	\\
            WordPress Android & 58,188 & 54,555 & 93.8\% & 3,633 & 6.2\% & 53,789 & 92.4\% & 4,399 & 7.6\%	\\
            \hline
        \end{tabular}%
        }
        \begin{tablenotes}
            \item[]{}
        \end{tablenotes}
    \end{threeparttable}
\end{table*}

We see that the percentages of different metric values are between 0.8\% and 6.2\%. 
Considering the different results in locations of changes, ranging from 1.4\% to 7.6\%, we found that quite a few portions of the metric values are same even though the identified locations are different.

To further explore of the disagreements between \textit{Myers} and \textit{Histogram}, we calculated the number of commits influenced by the different number of code changes and the locations in the \texttt{diff} output of files.  
In each project, we counted the sum of files that have the same and different quantity and the position of lines inserted and removed from each commit across the project. 
A single commit may contain more than one modified file.
If a commit recorded at least one file having unequal changed lines of code either in their number or their location, we classified this commit as `different'. 
On the other hand, if all files in a commit had identical changed lines, we categorized the commit in the `same' class. 
In this process, we only notify the files that have an unequal number and location of the lines of code. 

\begin{table*}[ht]
    \caption{The number of commits that contain a different number and the position of added and deleted lines of code in a file}
    \label{tab:summarymetrics}
    \centering
    \begin{threeparttable}
        \begin{tabular}{lrrrrr}
            \hline\noalign{\smallskip}
            \multirow{2}{*}{Project} & \multirow{2}{*}{\#Commits} & \multicolumn{2}{c}{Metrics (NLA and NLD)} & \multicolumn{2}{c}{Locations of Changes}	\\
            \cline{3-6}
            \noalign{\smallskip}
            & & \#Different & \% & \#Different & \%	\\
            \noalign{\smallskip}\hline\noalign{\smallskip}
            Apache Storm & 9,317 & 395 & 4.2\% & 587 & 6.3\%	\\
            Butterknife & 836 & 34 & 4.1\% & 48 & 5.7\%	\\
            Crate & 8,646 & 707 & 8.2\% & 1,202 & 13.9\%	\\
            Hystrix & 2,106 & 87 & 4.1\% & 160 & 7.6\%	\\
            JabRef & 11,940 & 881 & 7.4\% & 1,317 & 11.0\%	\\
            jcabi-github & 2,521 & 42 & 1.7\% & 70 & 2.8\%	\\
            Openmicroscopy & 46,543 & 2,340 & 5.0\% & 3,674 & 7.9\%	\\
            Presto & 13,561 & 816 & 6.0\% & 1,423 & 10.5\%	\\
            RxAndroid & 461 & 11 & 2.4\% & 18 & 3.9\%	\\
            SpongeAPI & 2,479 & 128 & 5.2\% & 235 & 9.5\%	\\
            Spring Boot & 17,087 & 954 & 5.6\% & 1,447 & 8.5\%	\\
            Square OkHttp & 3,171 & 106 & 3.3\% & 181 & 5.7\%	\\
            Square Retrofit & 1,576 & 61 & 3.9\% & 90 & 5.7\%	\\
            WordPress Android & 30,295 & 2,108 & 7.0\% & 3,050 & 10.1\%	\\
            \hline
        \end{tabular}
    \end{threeparttable}
\end{table*}

Our results show that several changed files impacted by the changed lines have similar commits. We grouped the same commits from these several files that contain different changed lines of code into a single commit. 
We then summarized the percentage of commits that have a different number and position of the changed lines of code resulting from the usage of the \textit{Myers} and \textit{Histogram} algorithms in the \texttt{git diff} command as described in Table \ref{tab:summarymetrics}.

In general, our comparisons revealed that the data extraction using two \texttt{diff} algorithms in the command produced identical \texttt{diff} lists for most files in all commits. 
However, even though the output has been dominated by the same results for each file in a commit, the \texttt{diff} output from the \textit{Myers} and \textit{Histogram} recorded several files that have different added and deleted lines. 
These disagreements impacted the dissimilar number of commits that have files containing changed lines of code. 
The level of differences in the number of commits influenced by the amount of lines of code are adequately high, ranging from 1.7\% to 8.2\%, while the unequal location of lines affects the level of differences in the quantity of commits from 2.8\% to 13.9\%. 

\subsection{Summary}
The finding from the metrics comparison provides clear evidence that the use of multiforms of \texttt{diff} algorithms might differentiate the \texttt{diff} lists.
Since the metrics are insensitive to differences in change locations, the same values can be obtained even if identified change locations are different. 
However, we see that different metric values were obtained from 0.8\% to 6.2\% in the file-level and 1.7\% to 8.2\% in the commit-level. 
These differences can have impacts on studies using \texttt{diff}-related metrics.

\section{Comparison: SZZ Algorithm (RQ$_2$)}
\label{sec:szz}

\RqTwo

\subsection{SZZ Algorithm}
\label{sec:szzalgorithm}
\begin{figure}[]
    \center
    \includegraphics[width=0.7\textwidth]{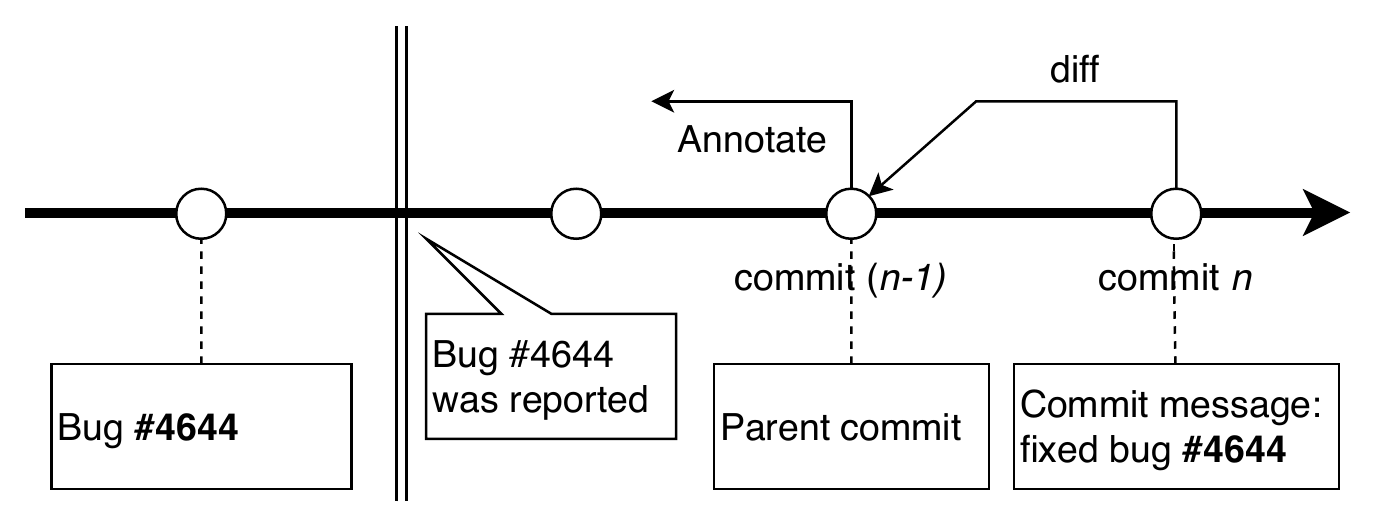}
    \caption{SZZ: Locating bug-introducing changes}
    \label{fig:szzalgorithm}
\end{figure}

The SZZ algorithm proposed by \cite{Sliwerski:2005:CIF:1082983.1083147} is an approach to identify bug-introducing changes. 
The SZZ uses a bug-tracking system (e.g. Bugzilla) as the reference to link archived versions of a software (e.g. CVS).
Figure~\ref{fig:szzalgorithm} depicts the basic idea of the SZZ algorithm. 

The SZZ algorithm first identifies bug-fixing commits by searching \textit{bug report identity numbers (bug ID)} in log messages, which have been written by developers when they fix bugs. 
The commit ID of this bug-fixing commit is subsequently used to track the previous commit (parent commit). 
The code changes are extracted by applying \texttt{diff} to find the differences between the older version of a file in the parent-commit and the newer version of the same file in the bug-fix commit.
The identified deleted lines are considered to be candidates of bug-related lines.
To identify bug-introducing commits, \texttt{cvs annotate} command is used to investigate when lines are added. Among the candidates of bug-related lines, lines that have been created before the bug reporting time are considered to be \textit{validated} bug-related lines. The commits that introduced those validated bug-related lines are identified as bug-introducing commits.

\subsection{Analysis Design}
\label{sec:analysisdesign_szz}

\begin{figure}[]
    \center
    \includegraphics[width=1\textwidth]{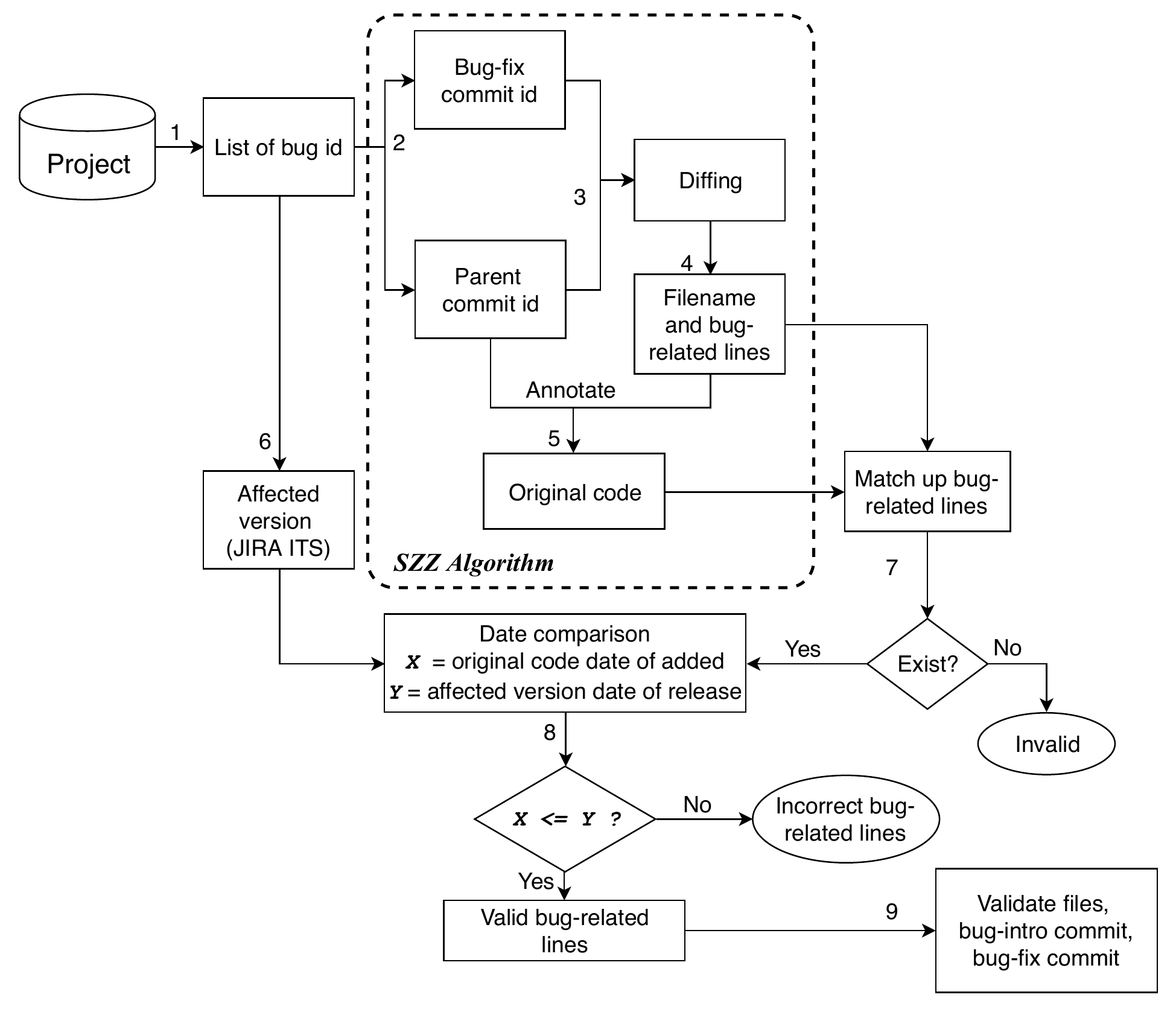}
    \caption{Overview of the validation process of bug-introducing commits}
    \label{fig:bugintro_validation_process}
\end{figure}

Figure~\ref{fig:bugintro_validation_process} describes the validation process of our analysis. 
For our empirical analysis, we studied 10 open source Apache projects used in the previous study~\citep{dacostaSZZ2017}, which is identified in our systematic mapping as a study utilizing Git for identifying bug introduction using the SZZ algorithm.
The descriptions of projects and the number of commits in the \textit{master} branches are shown in Table~\ref{tab:apacheproject}. 
We analyzed the impact of using different \texttt{diff} algorithms on the original SZZ algorithm.
We studied the disagreement between the \textit{Myers} and \textit{Histogram} in the results of the SZZ algorithm based on \texttt{diff}.

\begin{table*}[]
    \caption{Overview of the 10 studied Apache projects}
    \label{tab:apacheproject}
    \begin{threeparttable}
        \begin{tabularx}{\linewidth}{lp{6.7cm}r}
            \hline\noalign{\smallskip}
            Project Name & Description & \# Commits \\
            \noalign{\smallskip}\hline\noalign{\smallskip}
            ActiveMQ & Message broker and Java Message Service client & 9,962 \\
            Camel & A framework to create routing and mediation rules in various domain specific languages & 32,124 \\
            Derby & Relational database implemented in Java & 8,184 \\
            Geronimo & Open source application server compatible with Java EE & 13,137 \\
            Hadoop Common & Collection of common utilities and libraries that support other Hadoop modules & 10,509 \\
            HBase & A distributed big data store for the Hadoop database & 15,091 \\
            Mahout & A library to generate the implementation of distributed or scalable machine learning algorithms & 3,959 \\
            OpenJPA & The implementation of the Java Persistence API & 4,864 \\
            Pig	& High-level mechanism for the parallel programming of MapReduce on Hadoop & 3,154 \\
            Tuscany & An open source to develop applications based on SCA standard & 16,253 \\
            \hline
        \end{tabularx}
    \end{threeparttable}
\end{table*}

First, \textit{bug report IDs} in the commit messages are searched with specific keywords (i.e. ``bug'', ``fix'', ``defect'', and ``patch''~\citep{Sliwerski:2005:CIF:1082983.1083147}), then the identified commits are marked as candidates of bug-fixing commits. 
In each candidate bug-fixing commit, we focus on the modified files.
The two \texttt{diff} algorithms are used to identify deleted lines using the command: \texttt{git diff -w --ignore-blank-lines --diff-algorithm=<algorithm> <parent commit\\ ID> <bug-fix candidate commit ID> -- <filename>}.
By fetching files in the parent commit ID, we subsequently applied the \texttt{git blame} command (similar to \texttt{cvs annotate}) to locate the origin of the deleted lines.
Those deleted lines are considered to be candidates of bug-related lines.

Similar to the procedure of \cite{dacostaSZZ2017}, the next step is to find the \textit{affected software versions} of a bug. 
We extract bug reports and their \textit{affected versions} from the JIRA issue tracking system\footnote{\label{jiraitsweb}\url{https://www.atlassian.com/software/jira} (April 2018)}.
If a single bug ID affects more than one version, the earliest version is chosen since the SZZ algorithm targets the initial appearance of a bug. 
From the collection of \textit{affected-versions}, we compare the dates of the introduction of the candidates of bug-related lines with the release dates of the versions. 
If the release dates of the affected versions are later than the dates of the introduction of the candidates of bug-related lines, we classified them as \textit{valid bug-related lines}; otherwise, we classified them as \textit{invalid}. 

With these sets of valid bug-related lines, we validate bug-introducing commits, bug-related files and bug-fixing commits. 
The validation processes are performed in the opposite direction with the above procedure. 
A valid bug-introducing commit is a commit that initially adds valid bug-related lines. 
Files containing bug-related lines are considered to be valid bug-related files.
From the candidates of bug-fixing commits, if there is at least one valid associated bug-introducing commit, we consider the candidate bug-fixing commit to be valid, otherwise invalid.

\subsection{Results}
\label{sec:result_szz}

\begin{table*}[]
    \caption{Summary of valid bug-related lines, valid files, valid bug-introducing commits, and valid bug-fix commits resulting from \textit{Myers} and \textit{Histogram}}
    \label{tab:summaryvalidszz}
    \center
    \begin{threeparttable}
        \resizebox{\columnwidth}{!}{%
        \begin{tabular}{lrrrrrrrr}
            \hline\noalign{\smallskip}
            \multirow{2}{*}{Project}
            	& \multicolumn{2}{p{1.7cm}}{\#valid bug-related lines} & \multicolumn{2}{c}{\#valid files} & \multicolumn{2}{p{1.8cm}}{\#valid bug-intro commits} & \multicolumn{2}{p{1.6cm}}{\#valid bug-fix commits}	\\
            \cline{2-9}
            	& M & H & M & H & M & H & M & H	\\
            \hline
            ActiveMQ & 10,671 & 10,846 & 1,566 & 1,565 & 1,015 & 1,016 & 614 & 613\\
            Camel &	12,525 & 12,626 & 2,377 & 2,374 & 1,514 & 1,516 & 716 & 712	\\
            Derby &	130,861 & 131,031 & 4,372 & 4,373 & 1,178 & 1,180 & 1,038 & 1,039	\\
            Geronimo & 29,543 & 29,743 & 2,448 & 2,448 & 1,282 & 1,277 & 462 & 462	\\
            Hadoop Common & 15,053 & 15,285 & 805 & 805 & 546 & 550 & 318 & 318 \\
            HBase & 37,558 & 37,291 & 2,083 & 2,079 & 1,480 & 1,481 & 669 & 668	\\
            Mahout & 1,542 & 1,548 & 182 & 182 & 145 & 144 & 44 & 44	\\
            OpenJPA & 5,160 & 5,204 & 794 & 794 & 370 & 370 & 365 & 366	\\
            Pig & 1,789 & 1,787 & 205 & 206 & 187 & 187 & 80 & 80	\\
            Tuscany & 750 & 781 & 46 & 46 & 34 & 36 & 16 & 16	\\
            \hline
        \end{tabular}%
        }
        \begin{tablenotes}
            \item $M = Myers$
            \item $H = Histogram$
        \end{tablenotes}
    \end{threeparttable}
\end{table*}

Table \ref{tab:summaryvalidszz} presents the outputs of the \textit{Myers} and \textit{Histogram} algorithms in the number of valid bug-related lines, files, bug-introducing commits, and bug-fix commits. 
Two algorithms produced a different number of valid bug-related lines in all 10 projects, which then led to the different number of files, bug-introducing commits, and bug-fix commits.

Similar to the analysis of metrics in Section~\ref{sec:metric}, differences in the quantities of changes are relatively small or the same for some projects, because of the insensitivity of change locations.

Since investigating the locations of bug introduction is also important, we perform a comparison of files that have the same and different locations of bug-related lines.
Table~\ref{tab:numberoffileineachcommit} shows this result. 
It can be seen that the total number of files that have a different location of the changed code is high in each project, ranging from 2.4\% to 6.6\%. 
This means that some files can contain suspicious bug-related lines, only because of different algorithms.

\begin{table*}[t]
    \caption{Total number of files that have the same and different positions of valid bug-related lines in all valid bug-fix commits}
    \label{tab:numberoffileineachcommit}
    \centering
    \begin{threeparttable}
        \begin{tabular}{lrrrr}
            \hline\noalign{\smallskip}
            Project & \#Same & \#Different & \% & Total	\\
            \noalign{\smallskip}\hline\noalign{\smallskip}
            ActiveMQ & 1,464 & 103 & 6.6\% & 1,567	\\
            Camel & 2,315 & 63 & 2.7\% & 2,378	\\
            Derby & 4,198 & 175 & 4.0\% & 4,373	\\
            Geronimo & 2,318 & 130 & 5.3\% & 2,448	\\
            Hadoop Common & 777 & 28 & 3.5\% & 805	\\
            HBase & 1,973 & 110 & 5.3\% & 2,083	\\
            Mahout & 171 & 11 & 6.0\% & 182	\\
            OpenJPA & 764 & 31 & 3.9\% & 795	\\
            Pig & 201 & 5 & 2.4\% & 206	\\
            Tuscany & 43 & 3 & 6.5\% & 46	\\
            \hline
        \end{tabular}
        \begin{tablenotes}
            \item[]{}
        \end{tablenotes}
    \end{threeparttable}
\end{table*}

\pgfplotsset{compat=1.14}
\definecolor{diffcluster}{HTML}{BEBEBE}
\definecolor{samecluster}{HTML}{EDEDED}

\pgfplotstableread[col sep=comma,header=true]{
Project, 1, 2
ActiveMQ, 12.68, 87.32
Camel, 6.28, 93.72
Derby, 10.30, 89.70
Geronimo, 11.90, 88.10
Hadoop Common, 5.97, 94.03
HBase, 13.30, 86.70
Mahout, 11.36, 88.64
OpenJPA, 6.28, 93.72
Pig, 6.25, 93.75
Tuscany, 12.50, 87.50
}\data

\pgfplotstablecreatecol[
 create col/expr={
    \thisrow{1} + \thisrow{2}
 }
]{sum}{\data}

\pgfplotsset{
  percentage plot/.style={
    point meta=explicit,
    every node near coord/.append style={
      font=\tiny,
      color=black,
    },
    nodes near coords={
      \pgfmathtruncatemacro\iszero{\originalvalue==0}
      \ifnum\iszero=0
      \pgfmathprintnumber[fixed,fixed zerofill,precision=1]{\pgfplotspointmeta}
      \fi
    },
    yticklabel=\pgfmathprintnumber{\tick}\,$\%$,
    ymin=0,
    ymax=100.01, 
    visualization depends on={y \as \originalvalue},
    enlarge x limits={abs=6mm}
  },
  percentage series/.style={
    table/x expr=\coordindex, 
    table/y expr=(\thisrow{#1}/\thisrow{sum}*100),
    table/meta=#1
    }
}

\begin{figure}[]
    \center
    \begin{tikzpicture}
    \begin{axis}[
        ybar stacked,
        height=6cm,
        width=11cm,
        percentage plot,
        bar width=0.7cm, 
        xticklabels from table={\data}{Project}, 
        xtick=data,
        x tick label style={
          rotate=45,
          anchor=east, 
          xshift=-1.5mm, yshift=-2mm
        },
        legend style={
          at={(0.5,-0.7)},
          anchor=south,
          legend columns=-1
          },
    ]
    
        \addplot [fill=diffcluster]  table[percentage series=1] {\data};
        \addplot [fill=samecluster]       table[percentage series=2] {\data};
    
        \addplot [forget plot,nodes near coords align=above] table[x expr=\coordindex,y expr=0.0001,meta=sum]{\data};
    
        \legend{\strut Different location, \strut Same location}
    \end{axis}
    \end{tikzpicture}
    \caption{The percentage of valid bug-fixing commits that have the same and different positions of valid bug-related lines}
    \label{fig:percentageofvalidbugfixhavingdiffresult}
\end{figure}
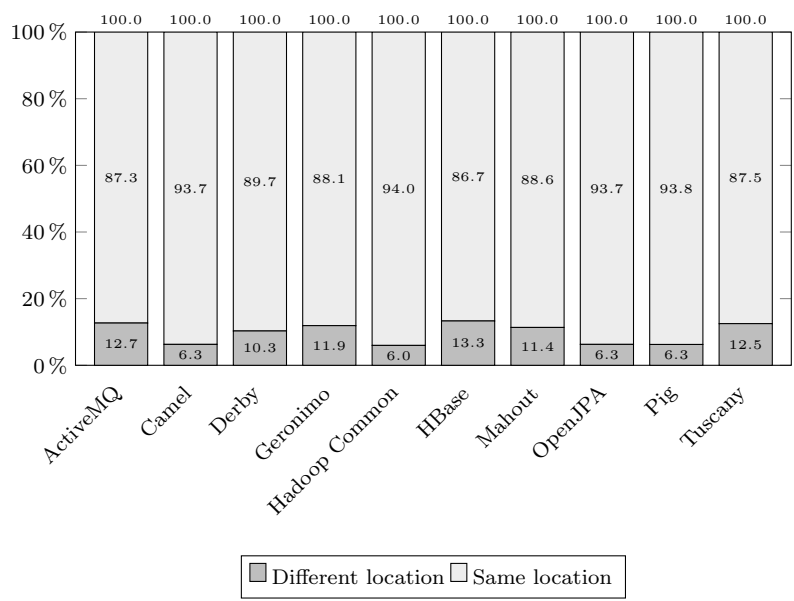

Bringing these data into further analysis, we then summarized the number of valid bug-fixing commits. 
As shown in Figure \ref{fig:percentageofvalidbugfixhavingdiffresult}, all studied projects have a different number of valid bug-fixing commits caused by the different positions of valid bug-related lines resulting from the \textit{Myers} and \textit{Histogram}. 
The percentage of the different results are between 6.0\% and 13.3\%, or 9.7\% on average. 
This analysis found evidence that nearly 10\% of bug-fixing commits do not guarantee success in locating bug-introducing changes since some deleted lines that were suspected as the candidate bug-introducing changes are different if we applied different \texttt{diff} algorithms in the \texttt{git diff} command. 
This is because a valid bug-related line in a file has the possibility of being identified by a particular \texttt{diff} algorithm, but it remains undetected while using the other \texttt{diff} algorithms.

\subsection{Summary}

The results from the SZZ algorithm confirm that different \texttt{diff} algorithms possibly generate different results, from 6.0\% and 13.3\% in the total of the identified bug-fix commits.
The \textit{Myers} and \textit{Histogram} sometimes produced a different number and location of the deleted lines (bug-related lines) in several files.
These differences certainly affect the number of disagreement files that have the bug-related lines, the amount of bug-introducing commits, and the bug-fixing commits that actually have the bug-contained files.
Therefore, the comparison result indicates that several prior studies that had used the SZZ algorithm to locate bugs have the possibility of producing inaccurate analyses.

\section{Comparison: Patches (RQ$_3$)}
\label{sec:patch}

\RqThree

\subsection{Analysis Design}
\label{sec:analysisdesign_patches}

From the previous two comparisons, we showed that different \texttt{diff} algorithms can have different results of metrics collection and bug-introduction identification (SZZ algorithm). 
Computationally, both \texttt{diff} algorithms are correct in textual differencing. 
However, the \texttt{diff} outputs are sometimes different due to different \texttt{diff} algorithms. 
The \texttt{diff} results might show different change region with a contiguous list of deleted and added lines that is called as a change hunk (Ray et al, 2015). 
We expect that a set of changing operations done by developers can be represented by change hunks. 
However, the identification of the change hunks can be inappropriate.
In our investigation, this issue could not be identified automatedly. 
Thus, we analyze the quality of \texttt{diff} manually.

To judge the quality of the \texttt{diff} algorithms, we define ``better'' if the algorithms meet our two criteria: (i) it detects the unmodified lines appropriately that should not be identified as changed lines, and (ii) it shows the changed lines more systematically~\citep{Kim:2013:6165314}. 
The sequences of the added and deleted lines of code are expected to be closer to what developers did to the code.
If the code elements change together, they are shown explicitly as group systematic changes or report their common structural characteristics.

For this analysis, we used the same dataset that had been used in Section~\ref{sec:metric} and Section~\ref{sec:szz}, shown in Table \ref{tab:no_of_population}. 
From the CI-Java projects, we considered all modified files in all commit IDs to be targeted, while of the Apache projects, files changed in all bug-fix commit candidates are targeted. 
We applied the same command as the other two comparisons: \texttt{git diff -w --ignore-blank-\\lines --diff-algorithm=<algorithm> <parent commit ID> <commit ID> -- <filename>} to generate the \texttt{diff} output from \textit{Myers} and \textit{Histogram}.
In each project of the first group, we analyzed the files that have different locations of the inserted and removed lines from the execution of the two \texttt{diff} strategies. 
While in the second group, only the files that have a different location of the deleted lines were analyzed.

\begin{table*}[]
    \centering
    \caption{Targeted files that have different locations in identified lines with two \texttt{diff} algorithms}
    \label{tab:no_of_population}
    \begin{tabular}{llr}
        \hline\noalign{\smallskip}
        Project Group & Type of identified line & \#Files \\
        \noalign{\smallskip}\hline\noalign{\smallskip}
        CI-Java Projects (Section~\ref{sec:metric}) & Added and deleted lines & 20,535    \\
        Apache Projects (Section~\ref{sec:szz}) & Deleted lines & 1,055  \\
        \noalign{\smallskip}\hline\noalign{\smallskip}
        Total & & 21,590  \\
        \noalign{\smallskip}\hline\noalign{\smallskip}
    \end{tabular}
\end{table*}

We divided the comparison into two categories: (i) \textit{in-code diff} and (ii) \textit{in-non-code diff}. 
The first category of \textit{diff} means the different \texttt{diff} lists generated by both algorithms are lines of code or a block of code in a source code file. 
Otherwise, the second \textit{diff} implies the disagreement between these two algorithms are other than a line of code, for example a change of comments, or a change in a non-code file, such as a modification in a text file.

Qualitative analysis between the two \texttt{diff} algorithms was performed manually by the first two authors in multiple steps. 
Initially, the first author made a list of all files from the two project groups. 
From this list, the sample size of files was counted using the tool provided in a survey system\footnote{\label{web:surveycalc}\url{https://www.surveysystem.com/sscalc.htm}} to statistically represent sample from files in each project, so that the conclusions about the quality of the \texttt{diff} algorithm would generalize to all files in all projects with a confidence level of 95\% and a confidence interval of 5. 
As can be seen in Table \ref{tab:no_of_population}, the total number of files summarized from all project groups is 21,590. 
From this population, we selected random samples of 377 files.

In the second step, we conducted a manual comparison between two \texttt{diff} outputs produced by \textit{Myers} and \textit{Histogram} algorithms from all files in the sample.
The first two authors of this paper were involved to independently annotate the \texttt{diff} outputs that makes the result is expected to be more reliable. 
To specify the comparison result between two \texttt{diff} algorithms, we generated three categories as described in Table \ref{tab:desc_diff_result}.
We assign \textbf{Histogram} to the comparison results if the \texttt{diff} outputs produced by \textit{Histogram} algorithm show the unmodified lines more appropriately and provide better group systematic changes to show the lines were changed together compared with the \textit{Myers}.
If the results produced by \textit{Myers} provide more appropriate unchanged contexts and show the group changes more systematically compared with the \textit{Histogram}'s \texttt{diff}, we labeled them as \textbf{Myers}.
While if the \texttt{diff} outputs produced by one algorithm are not better than the other, then we mark them the \textbf{Same}.
The comparison results between two authors from 377 files were subsequently computed to find the kappa agreement\footnote{\label{formula:kappa_agreement}\url{http://justusrandolph.net/kappa/}}. 
We obtained 70.82\%, which is categorized into `substantial agreement'~\citep{Viera:kappa:2005}. 
This means, the statistic result of our manual study is acceptable.

\begin{table*}[]
    \centering
    \caption{Description of the \texttt{diff} assessment}
    \label{tab:desc_diff_result}
    \begin{tabular}{lp{9.1cm}}
        \hline\noalign{\smallskip}
        Result & Condition    \\
        \noalign{\smallskip}\hline\noalign{\smallskip}
        Histogram & The output of \textit{Histogram} is better. \\
        Myers & The output of \textit{Myers} is better. \\
        Same & Both outputs are same level. One algorithm is not better than the other. \\
        \hline
    \end{tabular}
\end{table*}

\subsection{Results}
\label{sec:result_patches}
Table \ref{tab:freq_diffcomparison} shows how well both \texttt{diff} algorithms work in presenting the changes of code. 
It can be seen that \textit{Histogram} outnumbered the other results in the \textit{in-Code diff} category, which emphasizes that this algorithm is substantially better to differentiate the changes of code specifically.

\begin{figure}
    \centering
    \subfloat[\texttt{diff} output using \textit{Myers}\label{fig:diff_list_histogrambetter_a}]{
        \includegraphics[width=1\textwidth]{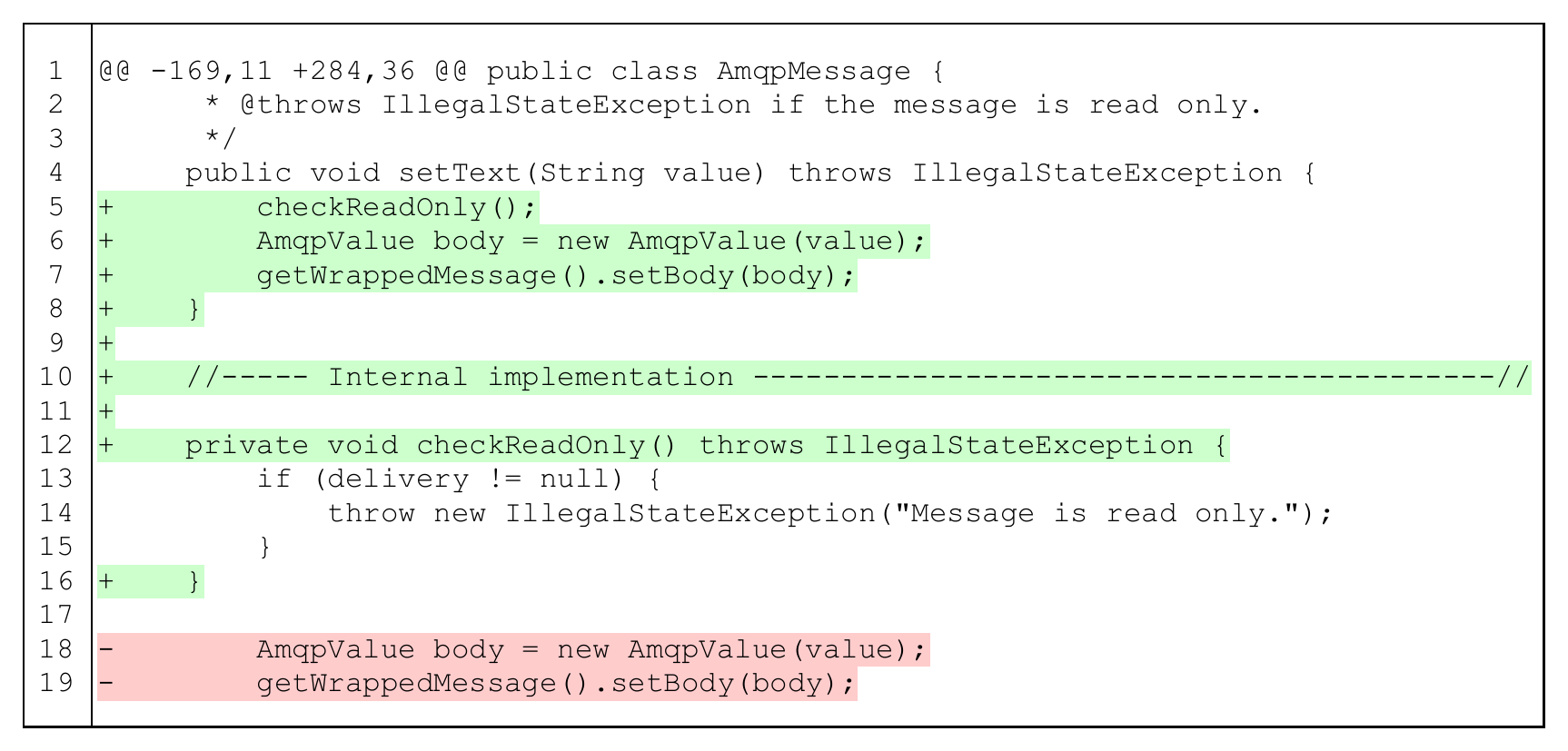}
    }
    \hfill
    \subfloat[\texttt{diff} output using \textit{Histogram}\label{fig:diff_list_histogrambetter_b}]{
        \includegraphics[width=1\textwidth]{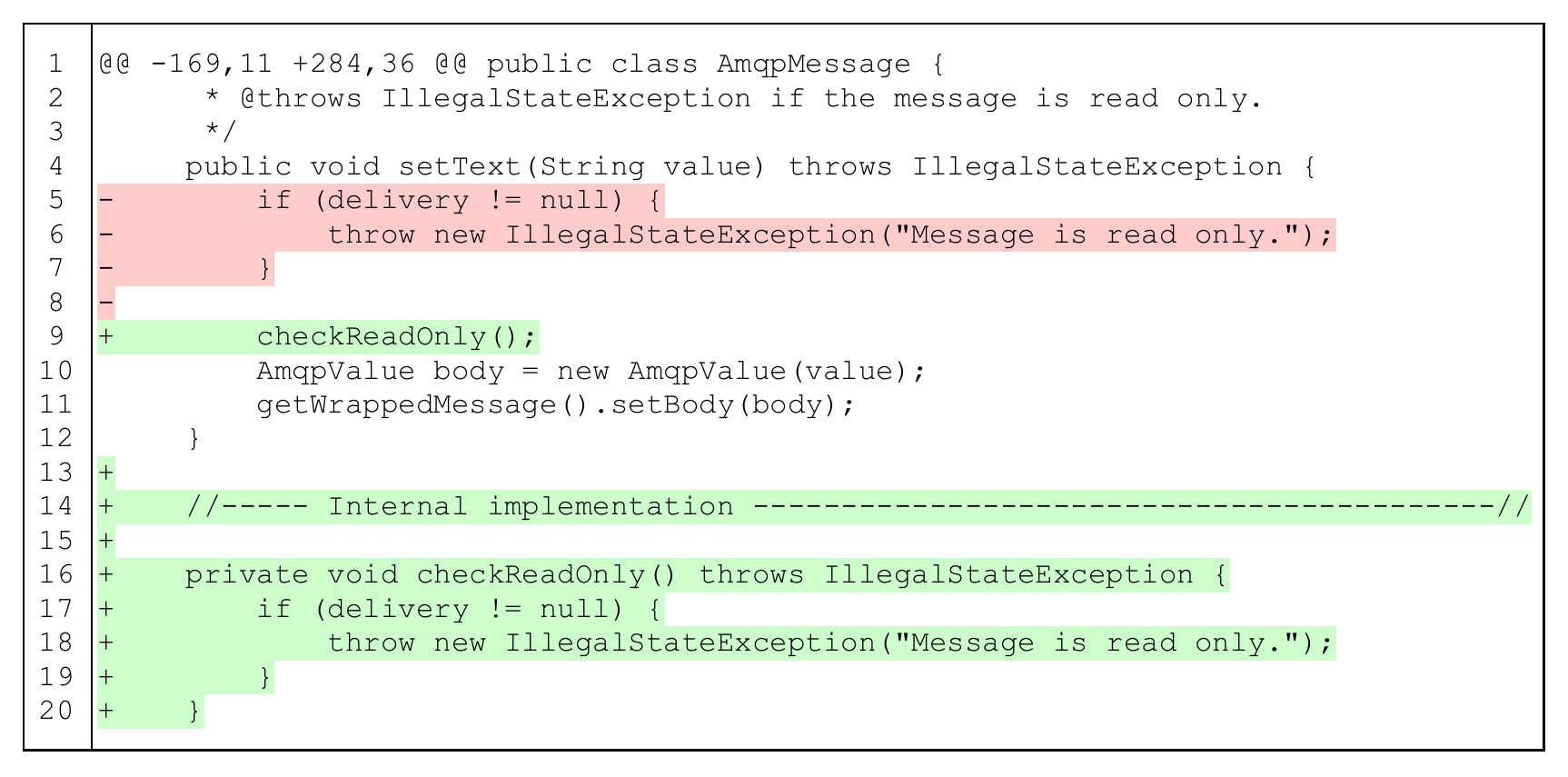}
    }
    \caption{Example of \texttt{diff} outputs generated by \textit{Myers} and \textit{Histogram} in extracting the code changes}
    \label{fig:diff_list_histogrambetter}
\end{figure}

Figure \ref{fig:diff_list_histogrambetter} shows how the \textit{Histogram} algorithm provides better output of code changes compared with the \textit{Myers}.
We extracted the \texttt{diff} from the file \texttt{AmqpMessage.java}\footnote{\label{web:amqpmessage}activemq-amqp/src/test/java/org/apache/activemq/transport/amqp/client/ AmqpMessage.java -- \url{https://github.com/apache/activemq/commit/f56ea45e58a17fa3aad46cbe8fc605ef4ffdbc81#diff-5296b90814217d75e272e14834a09dca}} in commit \texttt{f56ea45e5} from the project of \textit{ActiveMQ}. 
It is true that none of the algorithms are incorrect in describing changes. 
However, the \textit{Histogram} algorithm provides a reasonable diff output better describing human change intention, as the \textit{if}-statement is moved to a new method and a new method call is added.
While from the result of \textit{Myers}, it is not clear how developer changed the code. Lines that have not modified were identified as removed from the original positions (line 18 and 19) and added to the new positions (line 6 and 7).

\begin{table}[]
    \centering
    \caption{Frequency of comparison result in the sample data}
    \label{tab:freq_diffcomparison}
    \begin{tabular}{l|rr|rr}
        \noalign{\smallskip}\hline\noalign{\smallskip}
        \multicolumn{1}{l|}{Result} & \multicolumn{2}{c|}{\textit{in-Code diff}} & \multicolumn{2}{c}{\textit{in-Non-Code diff}} \\
        \noalign{\smallskip}\hline\noalign{\smallskip}
        Histogram & 152 & (62.6\%) & 18 & (13.4\%)   \\
        Myers & 41 & (16.9\%) & 20 & (14.9\%)    \\
        Same & 50 & (20.6\%) & 96 & (71.6\%) \\
        \noalign{\smallskip}\hline\noalign{\smallskip}
        Sum & 243 & (100\%) & 134 & (100\%)   \\
        \noalign{\smallskip}\hline
    \end{tabular}
\end{table}

\begin{figure}
    \centering
    \subfloat[\texttt{diff} output of comment using \textit{Myers}\label{fig:diff_noncode_myershistogram_a}]{
        \includegraphics[width=1\textwidth]{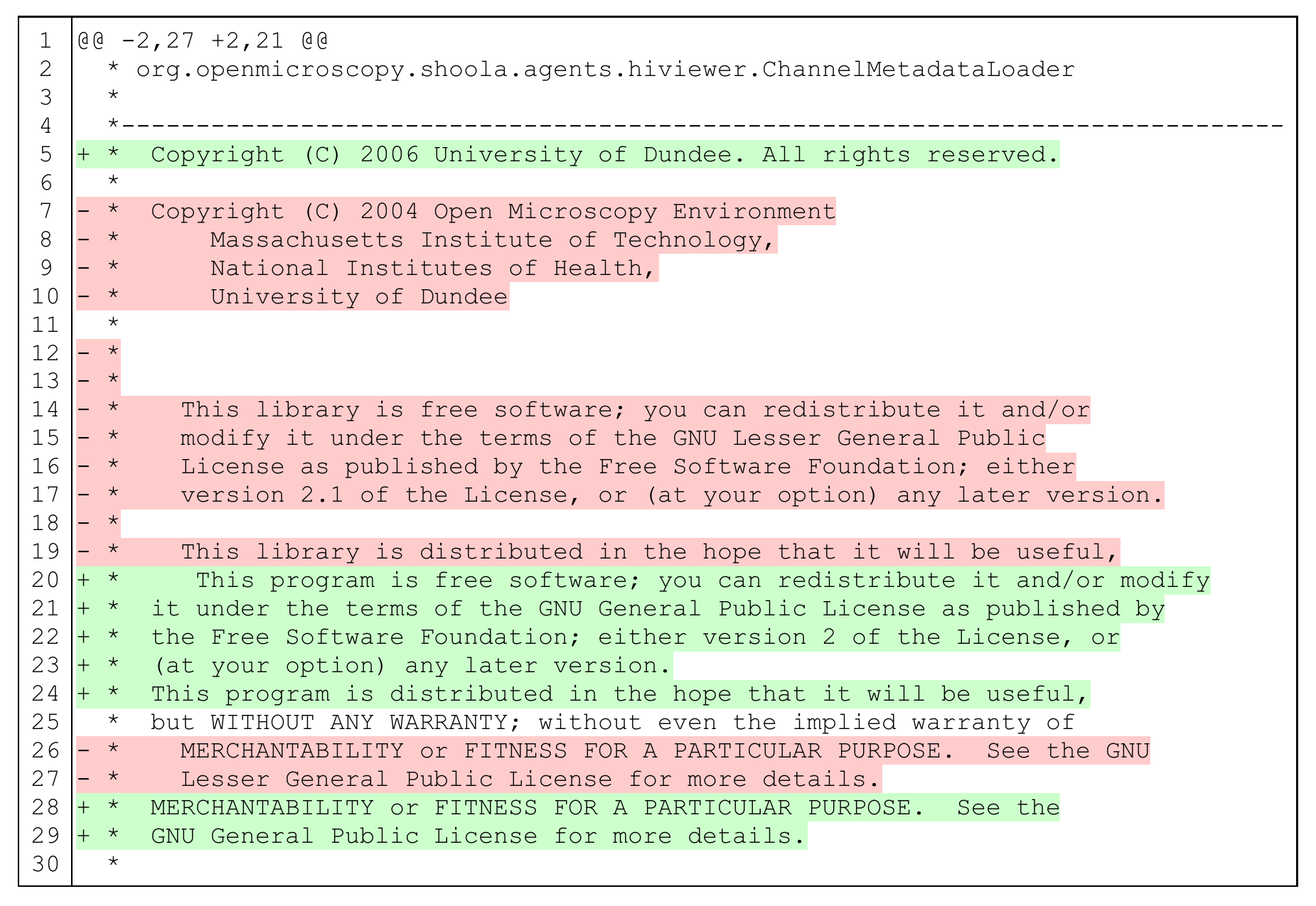}
    }
    \hfill
    \subfloat[\texttt{diff} output of comment using \textit{Histogram}\label{fig:diff_noncode_myershistogram_b}]{
        \includegraphics[width=1\textwidth]{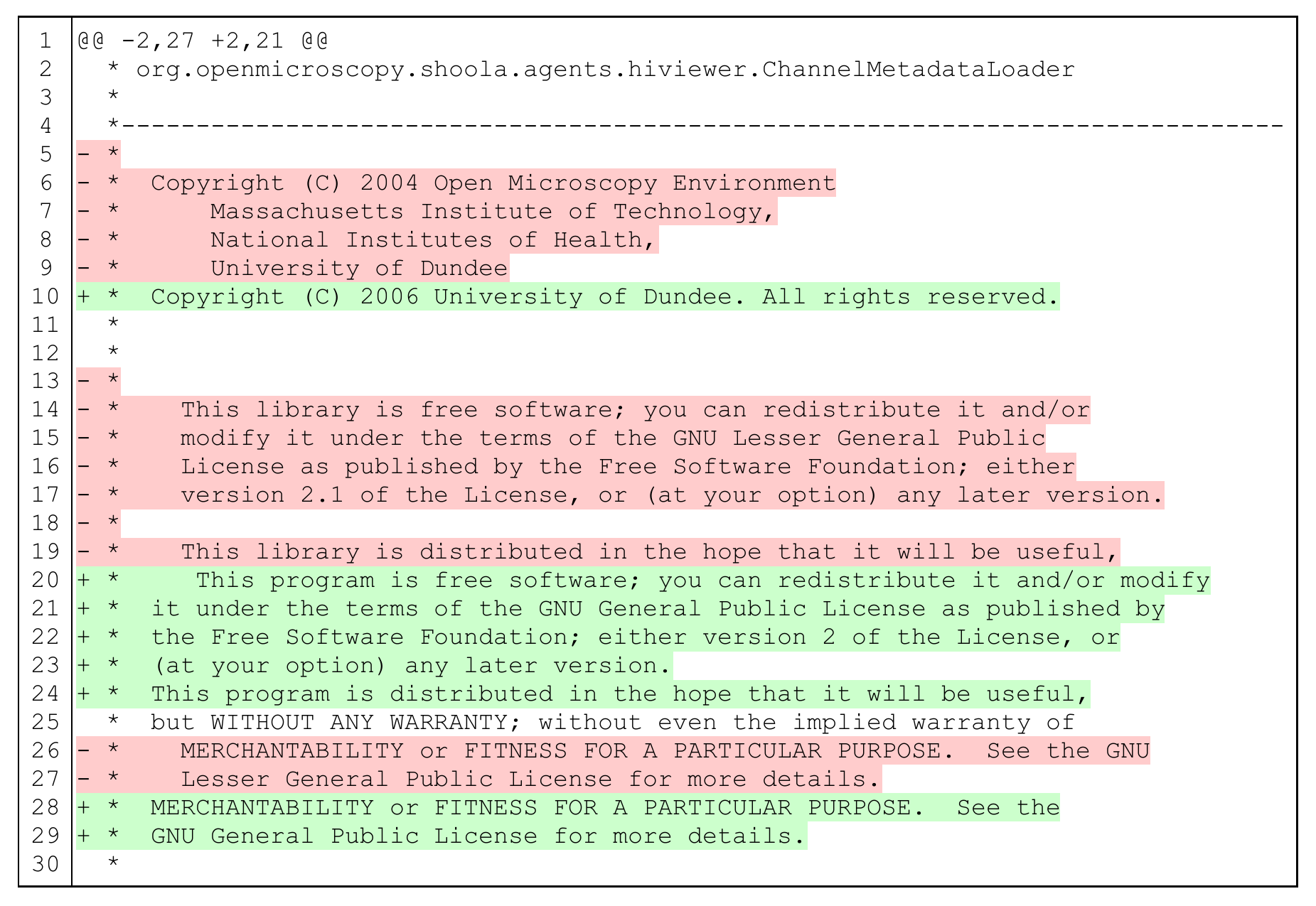}
    }
    \caption{Example of \texttt{diff} lists generated by \textit{Myers} and \textit{Histogram} in extracting the non-code changes}
    \label{fig:diff_noncode_myershistogram}
\end{figure}

This manual investigation also highlighted that the \textit{Myers} and \textit{Histogram} algorithms have almost the same ability to extract the \texttt{diff}s from non-code changes. 
As shown in Table \ref{tab:freq_diffcomparison}, their percentages are nearly equals in the \textit{in-Non-Code diff} (13.4\% files are better using the \textit{Histogram} and 14.9\% files are preferable using the \textit{Myers}). 
This is even strengthened by the high percentage of both \texttt{diff} algorithms' application that resulted in the same quality for the same files (see the example in Figure \ref{fig:diff_noncode_myershistogram}), which reached 71.6\%. 
This quantification reveals that we can use any of these algorithms to produce the \texttt{diff} from non-code changes. 
As shown in Figure \ref{fig:diff_noncode_myershistogram}, both \texttt{diff} algorithms worked well to reveal the comment changes from file \texttt{ChannelMetadataLoader.java}\footnote{\label{web:channelmetadataloader}SRC/org/openmicroscopy/shoola/agents/hiviewer/ChannelMetadataLoader.java -- \url{https://github.com/openmicroscopy/openmicroscopy/commit/e5924527fa467e117337b53a769503b6cc48e43f#diff-94a26d7e256533ade03740d86aca1afe}} in commit \texttt{e5924527fa} of \textit{Openmicroscopy} project since both lists are readable and understandable. 
The only differences between the two lists are the position of the initial added line and the matched line after the first inserted one. 
However, these disagreements did not change our interpretation about the modifications that occurred.

\subsection{Summary}

Due to the different procedures between \textit{Myers} and \textit{Histogram} in identifying the changed lines of code, they possibly generated different \texttt{diff} results.
Our manual comparison found that their differences were the number of the changes, the order of the changed lines, or even the detected added and deleted code.
They certainly affect the readability of the \texttt{diff} outputs, in other words, the quality of the \texttt{diff} results produced by the two \texttt{diff} algorithms were different.
Importantly, our results provide evidence that \textit{Histogram} frequently produced better \texttt{diff} results compared to \textit{Myers} in extracting the differences in source code.

\section{Discussions}
\label{sec:discussion}
\subsection{Implication and Recommendation}
In this paper, we present a description of the impact of different \texttt{diff}s on the results of a study. 
In the example shown in Figure \ref{fig:diff_list_histogrambetter}, we can see both algorithms identify the changed lines of code from line \#169. 
Nevertheless, there are several differences in the identified changed lines shown in both \texttt{diff} outputs.

The first difference is the number of the changed lines. 
From Figure \ref{fig:diff_list_histogrambetter}, we can see that the quantity of the detected changed lines are unequal. 
There are 11 changed lines discovered by the \textit{Myers}, while the \textit{Histogram} found 13 lines.
In a study that aims to collect metrics from the code changes, considering different \texttt{diff} algorithms is important since it has an impact on the number of changes.

In software quality analysis, one key factor of process metrics used to measure the changes is the number of modified lines (NLA and NLD).
For example, a work undertaken by~\cite{Gousios:2008:MDC:1370750.1370781} which proposed an approach to measure a software developer's contribution using \texttt{diff} records to compute the number of changed lines in a file.
This quantity of the changed lines was then used to calculate the commit size of all affected files. 
Based on our metrics comparison, we found that 1.7\% to 8.2\% commits have different NLA and NLD due to different \texttt{diff} algorithms application. 
While our manual investigation shows that more than 60\% \texttt{diff} outputs are better to extract using \textit{Histogram}.
Thus, if this study attempts to apply \textit{Histogram}, it might affect around 1\% to 4\% different commit size. 
As a result, this will impact the measurement of software developer's contribution as well.
Another study related to metrics analysis was conducted by~\cite{Rausch:2017:EAB:3104188.3104231}. 
The authors investigated the complexity of changes that can impact software quality.
The findings support that higher median values of NLA and NLD lead to an increase in build failures.
The study also found that the high mean values of the number of modified files correlates to the failed builds.
Based on the result from our metrics analysis, we found 0.8\% to 6.2\% files have different NLA and NLD.
Therefore, if \textit{Histogram} is applied in this study, this will influence around 0.5\% to 3.5\% of the modified files that correlates to the failed builds.

The second difference is the position of the changed lines.
Figure \ref{fig:diff_list_histogrambetter} shows that the two \texttt{diff} algorithms detect the deleted lines differently. 
The \textit{Myers} identifies one line of `Assignment' and one line of a `Method' call, whereas the \textit{Histogram} specifies a block of `\textit{if} condition'.
Related to SZZ application, both \texttt{diff} algorithms produce different deleted lines that are considered as the candidate of bug-introducing changes.
Thus, the identified bug-related lines might be invalid due to different \texttt{diff} algorithms application that can lead to the failure of bug-introducing changes identification.

A study undertaken by \cite{dacostaSZZ2017} investigates the output of five SZZ procedures in discovering the bug-introducing changes. 
The study on 10 Apache projects analyzed the validity of bug-introducing changes.
The validation process of bug-related lines used by the authors is similar to our study.
It compares the release dates of the earliest affected software versions of a bug with the dates of the introduction of the candidates of bug-related lines.
However, in our study, we enhanced the process to validate the other three parameters, that is, the bug-introducing commits that initially adds the valid bug-related lines, files containing valid bug-related lines, and bug-fixing commits that relates to valid bug-introducing commits. 
Our SZZ analysis shows that different \texttt{diff} algorithms application can have impact on the results of SZZ algorithm.
We found 2.4\% to 6.6\% valid files have different location of valid bug-related lines.
Since the \textit{Histogram} is better in more than 60\% \texttt{diff} outputs based on our manual analysis, therefore, if the study by \cite{dacostaSZZ2017} applies \textit{Histogram} in the \texttt{diff} command, around 1.5\% to 4\% files might have different valid bug-related lines in their study results.
The SZZ algorithm has also been studied by \cite{RODRIGUEZPEREZ2018164}. 
The authors conducted a literature review of published articles that focus on the SZZ algorithm's functionality and its ability to be imitated.
The similarity of this study to ours is investigating the changing impact due to the modification of SZZ algorithm. 
However, the study focus on the usability of the changing SZZ in the academic paper over time while our study analyze the impact of different \texttt{diff} algorithms application in the SZZ to study results.
Without considering the version of SZZ used in 187 previous studies collected by \cite{RODRIGUEZPEREZ2018164}, we understand that SZZ is a widespread and well-known algorithm over a 10-year period. 
This bug identification algorithm was commonly used to investigate commit size (26\% of the papers), line of code (15\% of the papers), number of changes (12\% of the papers), number of affected files (8\% of the papers), etc.
As described in our SZZ analysis, \texttt{diff} algorithms also have an impact on SZZ.
Thus, if the \textit{Histogram} is applied in those 187 prior studies, it might affect the results of studies.

Our investigations on metrics and SZZ application provide evidences that different \texttt{diff} algorithms application in \texttt{git} command can have an impact on a study result. 
It is also acknowledged that the \textit{Histogram} algorithm is substantially better than the \textit{Myers} to produce the changed lines of code.
Thus, we recommend to use the \textit{Histogram} in \texttt{git diff} command to extract the changes from source code.

\subsection{Threats to Validity}
\label{sec:threatstovalidity}

Threats to the \textit{construct validity} appear in the mapping study and the SZZ application. 
In our mapping study, we selected only the papers that specifically mention the \texttt{git} commands.
As a result, papers that had used \texttt{git} commands but do not mention it in the full text had been ignored, which can cause selection bias. Since different \texttt{diff} algorithms produce different results, we consider that papers should mention algorithm names of \texttt{diff} if the authors intentionally chose them.
In the SZZ application, we used a small number of keywords to detect commit messages that describe fixing bugs.
This limited our ability to extract all potential candidate bug-fixing commits.
Even so, the commits that should not be identified as bug-fixing commits were also possible to be collected as long as they included the keywords in their log messages.
However, since our focus is to investigate the level of differences of the \texttt{diff} lists produced by \textit{Myers} and \textit{Histogram}, the impact of the incorrect commits to the study result is small.
Another threat to the \textit{construct validity} is the definition of \textit{better} for the \texttt{diff} algorithm. 
We consider good quality of the algorithm based on our two criteria, while many could have been considered.
Different software engineering tasks may have different requirements for \texttt{diff} analysis.
However, since our focus is expecting to recover the changing operations from the \texttt{diff} outputs, the impact of this issue is not significant.

Threats to the \textit{external validity} emerge regarding the repository used in our experiments. 
Although we analyzed 24 OSS Java projects mined from Git repositories, we cannot generalize our study results to other open source projects nor industry.

To reduce the threats to \textit{reliability}, we make our dataset publicly available.
We provided lists of our collected files identified by the \textit{Myers} and \textit{Histogram} algorithms which were used in the three empirical analyses (see on GitHub\footnote{\dataset}).

\section{Conclusion}
\label{sec:conclusion}

To understand the impact of using different \texttt{diff} algorithms, \textit{Myers} and \textit{Histogram}, we first clarified applications of \texttt{diff} by conducting a systematic mapping of papers published between 2013 and 2017.
We then empirically analyzed the impact in three major applications:
(i) code churn metrics, (ii) SZZ algorithm, and (iii) patches extraction.

Our quantitative analyses has shown that the different \texttt{diff} algorithms can report different amount of changed lines, identify different change locations.
Our qualitative investigation revealed that \textit{Histogram} is better for describing code changes.
Since \texttt{diff} is the fundamental tool for various software engineering tasks, considering limitations and advantages of algorithms is important. 
Currently we recommend using the \texttt{Histogram} algorithm when analyzing code changes.



\end{document}